\documentclass[aps,reprint,amsmath,amssymb,pre,longbibliography,superscriptaddress]{revtex4-2}
\usepackage[utf8]{inputenc}
\usepackage{bm}
\usepackage{graphicx}
\usepackage{epstopdf}
\usepackage{hyperref}
\usepackage[hyphenbreaks]{breakurl}
 \usepackage{color}
\newcommand\beq{\begin{equation}}
\newcommand\eeq{\end{equation}}
\newcommand\beqa{\begin{eqnarray}}
\newcommand\eeqa{\end{eqnarray}}
\newcommand{\nn}{\nonumber\\}
\def\bal#1\eal{\begin{align}#1\end{align}}

\def\zero{{(0)}}
\def\one{{(1)}}
\def\two{{(2)}}

\newcommand{\py}{\text{PY}}

\newcommand\Nc{N_c}

\newcommand{\openonebis}{\mathsf{I}}

\newcommand{\tail}{\text{tail}}

\newcommand{\num}{\text{num}}
\newcommand{\talpha}{\alpha}
\newcommand{\tomega}{\omega}
\newcommand{\tdelta}{\delta}
\newcommand{\trho}{\rho}
\newcommand{\tA}{A}
\newcommand{\tr}{r}
\newcommand{\tk}{k}

\newcommand{\DD}{\mathcal{D}}

\begin{document}

\title{Structural properties of additive binary hard-sphere mixtures. III. Direct correlation functions}
\author{S{\l}awomir Pieprzyk}
\email{slawomir.pieprzyk@ifmpan.poznan.pl}
\affiliation{Institute of Molecular Physics, Polish Academy of Sciences, M. Smoluchowskiego
17, 60-179 Pozna\'n, Poland}
\author{Santos B. Yuste}
\email{santos@unex.es}
\homepage{http://www.unex.es/eweb/fisteor/santos/}
\author{Andr\'es Santos}
\email{andres@unex.es}
\homepage{http://www.unex.es/eweb/fisteor/andres/}
\affiliation{Departamento de F\'{\i}sica  and Instituto de Computaci\'on Cient\'{\i}fica Avanzada (ICCAEx), Universidad de
Extremadura, Badajoz, E-06006, Spain}
\author{Mariano L\'{o}pez de Haro}
\email{malopez@unam.mx}
\homepage{https://www.ier.unam.mx/academicos/mlh/}
\affiliation{Instituto de Energ\'{\i}as Renovables, Universidad Nacional Aut\'onoma de M\'exico (U.N.A.M.),
Temixco, Morelos 62580, M{e}xico}
\author{Arkadiusz C. Bra\'nka}
\email{branka@ifmpan.poznan.pl}
\affiliation{Institute of Molecular Physics, Polish Academy of Sciences, M. Smoluchowskiego
17, 60-179 Pozna\'n, Poland}

\begin{abstract}
An analysis of the direct correlation functions $c_{ij} (r)$ of binary additive hard-sphere mixtures of diameters $\sigma_s$ and $\sigma_b$ (where the subscripts $s$ and $b$ refer to the ``small'' and ``big'' spheres, respectively), as obtained with the rational-function approximation method and the WM scheme introduced in previous work [S.\ Pieprzyk \emph{et al.}, Phys.\ Rev.\ E {\bf 101}, 012117 (2020)], is performed.  The results indicate that the functions $c_{ss}(r<\sigma_s)$ and $c_{bb}(r<\sigma_b)$  in both approaches are monotonic and can be well represented by a low-order
polynomial, while the function $c_{sb}(r<\frac{1}{2}(\sigma_b+\sigma_s))$ is not monotonic and exhibits a well defined minimum near $r=\frac{1}{2}(\sigma_b-\sigma_s)$, whose properties are studied in detail. Additionally, we show that the second derivative $c_{sb}''(r)$ presents a jump discontinuity at $r=\frac{1}{2}(\sigma_b-\sigma_s)$ whose magnitude satisfies the same relationship with the contact values of the radial distribution function as in the Percus-Yevick theory.
\end{abstract}

\date{\today}

% insert suggested keywords - APS authors don't need to do this
%\keywords{}

%\maketitle must follow title, authors, abstract, \pacs, and \keywords
\maketitle

\section{Introduction}
\label{intro}

Systems composed of hard spheres (HSs) are important in the description of fluids, often playing the role of a generic or reference model system. They have become one of the most investigated off-lattice many-body physical systems, and considerable knowledge on them has been accumulated over decades. Nevertheless, the development of additional theoretical methods and current possibilities to perform effective simulations of large numbers of particles open up the opportunity to investigate in more depth or reveal new, hardly identified features of HS systems. This is especially important in the case of HS mixtures, which are obviously more complex than monocomponent systems.

In this paper, we continue with a series dealing with the behavior of the structural correlation functions of additive binary hard-sphere (BHS) mixtures. In the first paper \cite{PBYSH20},  we presented a method, referred to as the WM method,  which combines molecular dynamics (MD) simulation data, residue theorem analysis, and the Ornstein--Zernike (OZ) relations, allowing one to obtain an accurate representation of the structural correlation functions of this kind of mixtures. Both the above method and the so-called rational-function approximation (RFA), which turn out to be in very good agreement with each other, were employed in the same paper to test the direct correlation functions and to confirm the presence of a structural crossover for a particular mixture, namely, one with a fixed diameter ratio $q=0.648$ and a fixed total packing fraction  $\eta=0.5$ (which was the system analyzed previously theoretically and through experimental data by Statt \emph{et al.}\ \cite{SPTER16}). In the second paper \cite{PYSHB21}, we used the same methodology  to carry out a more thorough analysis of the role of the pole structure of the Fourier transforms of the total correlation functions $h_{ij}(r)$ of various BHS mixtures on the asymptotic behavior $r \to \infty$ of $h_{ij}(r)$, and its relation with structural crossovers in these functions. This allowed us, on the one hand, to confirm the power of our theoretical tool to study structural properties in BHS mixtures and, on the other hand and in the same vein as in Ref.\ \cite{GDER04},  to discuss a coarse-grained scenario that provides a fair picture of what goes on in the plane $\eta_s$ vs $\eta_b$ (where $\eta_s$  and $\eta_b$ are the partial packing fractions of the ``small'' and ``big'' spheres, respectively) when one varies the size ratio $q=\sigma_s/\sigma_b$ of the mixture (where $\sigma_s$  and $\sigma_b$ are the small and big diameters, respectively).  In the present third paper we continue with a further use of our theoretical tools and concentrate on the analysis of the direct correlation functions (DCFs) of BHS mixtures, which are some of the most hardly accessible and least studied structural properties of these systems.

In a simple fluid, the DCF $c(r)$ may be computed as the second derivative of the intrinsic free energy functional with respect to the number density $\rho$ \cite{HM13}, but it is usually
defined through  the OZ relation
\begin{equation}
 \label{OZ1}
  h(r_{12})=c(r_{12})+\rho \int d \mathbf{r}_3\, c(r_{13})h(r_{23}),
  \end{equation}
where $ h(r)=g(r)-1$ is the total correlation function, $g(r)$ being the radial distribution function. The subscripts ($1$, $2$, and $3$) denote the positions of three particles, where the separation between particles $i$ and $j$ is $r_{ij}=|\mathbf{r}_i-\mathbf{r}_j |$.
The function $c(r)$ represents that part of the total correlation function which results from the direct correlation between particles $1$ and $2$, and is also connected with the dimensionless isothermal compressibility of the fluid ($\chi$) via the exact relation
\begin{equation}
 \label{isocomp1}
  \chi^{-1}=1-\rho \tilde{c}(0)=1-\rho \, \int d \mathbf{r} \,c(r),
  \end{equation}
with $\tilde{c}(0)$ denoting the zero wave number value of the Fourier transform $\tilde{c}(k)$ of $c(r)$. Its importance in the theory of liquids may be judged from the following facts. First of all, if $c(r)$ is available, $h(r)$ (and hence the corresponding equation of state) may be readily obtained using the OZ relation. Such availability also allows one to get insight into how the presence of density fluctuations affects the free energy of the system. Further, knowledge of $c(r)$ in a homogeneous fluid may serve to develop approximate free energy density functionals for the inhomogeneous system.

Various aspects of the DCFs of hard-core systems have been reported in the literature. While the following description is certainly far from complete, we shall attempt to provide here an overview of the main developments. To our knowledge, the earliest analytic result for $c(r)$ in a three-dimensional system was reported by Wertheim \cite{W63} for a homogeneous HS fluid within the Percus-Yevick (PY) approximation. An exact result up to third order in density for this correlation function was obtained by Ashcroft and March \cite{AM67}. More recently, other approximate expressions for the DCFs of hard-core systems have been proposed. For hard-disk fluids, where there is no analytical solution in the PY theory, an expression for $c(r)$ was introduced by Ripoll and Tejero \cite{RT95}, who also considered a generalization valid for hard-core fluids in arbitrary $d$ dimensions. The DCF of two- and three-dimensional systems was also addressed by Guo and Riebel \cite{GR06}, who derived yet other approximate expressions for a monolayer of monodisperse hard disks and spheres. Due to the fact that in HS systems specification of the tail of the DCF is enough to derive the total correlation function $h(r)$ for all distances $r$, such a tail was analyzed by Henderson and Grundke \cite{HG75}, who introduced a parametrization of the tail to obtain an expression for the DCF of a HS fluid. A similar analysis of the tail of $c(r)$  was carried out by Katsov and Weeks \cite{KW00} for fluids whose molecules interact via a potential with a soft repulsive core of finite extent and a weaker and longer ranged tail.  Baus and Colot \cite{BC87} also derived an approximation of $c(r)$ for hard-core fluids in $d$ dimensions using rescaled virial expansions. While some of us \cite{YSH00a,TH07} derived the explicit expression for $c(r)$ for a HS fluid in the RFA, another two of us obtained an accurate representation of this DCF using the WM scheme \cite{PBH17}, and Fukudome \emph{et al.}\ \cite{FMI14} obtained an approximate $c(r)$ of the same system in  connection with scaled particle theory. Numerical simulation results for the function $c(r)$ of a HS fluid have been reported by Groot \emph{et al.}\ \cite{GEF87}. Analytical expressions for the DCFs in a multicomponent HS mixture were derived by Lebowitz \cite{L64} from the exact solution of the corresponding PY equation, while those of the RFA for the same system have been derived by some of us \cite{YSH00a,SYH20}. A sixth-order virial expansion was used by Dennison \emph{et al.}\ \cite{DMCA09} to obtain the DCF of a HS fluid that presents good agreement with simulation data. In a different vein, the DCFs of symmetric equimolar BHS mixtures  with negatively nonadditive diameters have been computed by Gazillo \cite{G88} in the PY approximation and those of the Widom--Rowlinson model by Fantoni and Pastore \cite{FP04} through Monte Carlo numerical simulations. Also using Monte Carlo data, Henderson \emph{et al.}\ \cite{HCD94} obtained DCFs for HSs near a large HS.\\
\indent Among the different systems considered within density-functional-theory approaches in which the DCFs are involved, the following few have been selected for this brief overview. Samborsky and Evans \cite{SE94} calculated the phase diagram of binary liquid crystal mixtures made of HSs and hard ellipsoids, while, by generalizing Rosenfeld's density functional theory for HS mixtures \cite{R89}, Charmoux and Perera \cite{CP96}  derived analytical approximations for the DCFs of molecular fluids and their mixtures. A simple weighted density approach for the one-particle correlation functions of the nonuniform system, requiring as input only the one- and two-particle DCFs of the corresponding uniform system, was used by Patra \cite{P99} to study the structure of BHS mixtures near a hard wall. Another simple weighted density approximation was considered by Zhou and Ruckenstein \cite{ZR00b} to derive DCFs of uniform fluids of all orders, finding that, in the case of uniform HS fluids, the third-order DCF was in satisfactory agreement with simulation data. In a similar path, Roth \emph{et al.}\ \cite{RELK02} developed a density functional for HS mixtures which has the same structure as the one of Rosenfeld's fundamental measure theory \cite{R89},  but also includes the Boubl\'ik-Mansoori-Carnahan-Starling-Leland (BMCSL) bulk equation of state \cite{B70,MCSL71}. By considering a generic free energy functional which requires the knowledge of the DCF of the homogeneous solvent (a quantity that may be extracted directly from MD simulations of the pure solvent), Ramirez \emph{et al.}\ \cite{RMB05} computed the DCFs of polar solvents. Moradi and Khordad \cite{MK06} used a formalism based on the work by Chamoux and Perera \cite{CP96} mentioned above to obtain the DCFs of binary mixtures of hard Gaussian overlap molecules, while Avazpour and Moradi \cite{AM07} combined the PY DCF and the one introduced by Roth \emph{et al.}\ \cite{RELK02} to obtain a  new expression for the DCF of HS fluids which they afterwards used to calculate the DCF of hard ellipsoidal fluids. With the aim of providing reference results for on-lattice density functional theories and related perturbation theories, Siderius and Gelb \cite{SG09} used both simulation results and theory to obtain thermodynamic and structural properties of on-lattice HS fluids. More recently, Lutsko \cite{L13} derived the DCF from the consistent fundamental-measure free energies \cite{S12c} for HS mixtures. Finally, Lin \emph{et al.}\ \cite{LOHHFK21}, on the basis of the fundamental-measure concept,  computed the DCF of a HS crystal and showed that it differs significantly from its liquid counterpart at coexistence.\\
\indent After this overview of the literature pertaining to the DCF, we turn to the subject with which this paper is mainly concerned. In the case of BHS  mixtures, there are three DCFs: $c_{bb}(r)$, $c_{ss}(r)$, and $c_{sb}(r)$. The functions $c_{ij}(r)$ are known to present a discontinuity at the contact distance $\sigma_{ij}=\frac{1}{2}(\sigma_i+\sigma_j)$, exhibiting the three of them an oscillatory fast decaying behavior for $r>\sigma_{ij}$. As we will discuss below, the behavior inside the core (i.e., $r<\sigma_{ij}$) is qualitatively different for the different DCFs. In particular, the functions $c_{bb}(r)$ and $c_{ss}(r)$ are monotonically increasing (concave) functions, similarly to what occurs with $c(r)$ in the monocomponent case. On the other hand, the cross DCF $c_{sb}(r)$ changes  very little (it is almost flat) up to a certain distance near $r=\lambda_{sb}\equiv\frac{1}{2}(\sigma_b-\sigma_s)$ and then increases considerably. The form of this latter function has not been systematically studied and it is usually thought to be a monotonic function with a constant value in the range $0<r<\lambda_{sb}$, as predicted by the PY theory \cite{L64,BH76}, which is the standard approximation used in the literature \cite{AM67,RT95,CP96,P99,ZR00b,FP04,AM07}.\\
\indent As pointed out later, we have found that, in fact, the function $c_{sb}$ is not a monotonically increasing function  for $r<\sigma_{ij}$ but presents a state-dependent minimum. One major aim of this paper is to reveal details of the DCFs with the focus on their core part, which represents a substantial part of the whole function. In passing, we will also establish that the RFA can predict well the features of the DCFs of additive BHS mixtures and, in particular, the nonmonotonic behavior of $c_{sb}(r)$ inside the core.\\
\indent The paper is organized as follows. In Sec.\ \ref{sec2},  we recall the explicit expressions of the Fourier transforms of the DCFs in terms of the Fourier transforms of the total correlation functions that follow from the OZ relation. In order to make the paper self-contained, in  Secs.\ \ref{WM} and \ref{secRFA}
we provide the explicit results for the Fourier transforms of the DCFs in BHS mixtures, as obtained with the WM scheme and the RFA, respectively. Section \ref{sec3}  profits from the previous derivation, allowing us to explicitly compute the DCFs of different mixtures. This is complemented with a comparison between the results of the WM method and the RFA predictions, as well as with the outcome of the PY theory and a subsequent discussion. The paper is closed in Sec.\ \ref{sec4} with some concluding remarks.

\section{Methods}
\subsection{The Ornstein--Zernike relation and the direct correlation functions}
\label{sec2}

The DCFs $c_{ij}({r})$ in a general $\Nc$-component mixture are defined through the OZ relation,
\begin{equation}
 \label{OZMulti}
  h_{ij}(r_{12})=c_{ij}(r_{12})+\rho\sum_{\ell=1}^{N_c} x_\ell \int d \mathbf{r}_3\, c_{i\ell}(r_{13})h_{\ell j}(r_{23}),
  \end{equation}
where  $\rho $ is the number density of the mixture  and $x_{i}=\rho _{i}/\rho $ is the mole fraction of species $i$ (where $\rho_i=N_i/V$ is the partial number density, $N_i$ and $V$ being the number of particles of species $i$ and  the volume of the system, respectively). In Fourier space, the OZ relation takes the following form \cite{S16}:
\begin{equation}
\label{Eq:OZ1}
\tilde{h}_{ij}(k)  = \tilde{c}_{ij}(k) +  \rho \sum_{\ell =1}^{\Nc} x_\ell\tilde{c}_{i\ell}({k}) \tilde{h}_{\ell j}({k}),
\end{equation}
where $\tilde{h}_{ij}(k)$ and $\tilde{c}_{ij}(k)$ denote the corresponding Fourier transforms of ${h}_{ij}(r)$ and ${c}_{ij}(r)$, $k$ being the wave number. They are given by
\begin{equation}
\label{FT}
\tilde{h}_{ij}({k})=4\pi \int_{0}^{\infty} dr\,r ^2 {h}_{ij}({r}) {{\sin{(\tk \tr)}} \over {\tk \tr}},
\end{equation}
with a similar expression for $\tilde{c}_{ij}({k})$.

We shall now restrict ourselves to the case of an additive BHS fluid mixture  in which the species of small spheres is labeled as species $s$ and the one of big spheres is labeled as species $b$. In this system,  the  hard core of the interaction between a sphere of species $i$ and a sphere of species $j$ ($i,j=s,b$) is given by  $\sigma_{ij}=\frac{1}{2}(\sigma _{i}+\sigma _{j})$, with the diameter of a sphere of species $i$ being $\sigma _{ii}=\sigma _{i}$. Let the size ratio be $q=\sigma_s/\sigma_b < 1$. In this instance, one can define the partial packing fractions $\eta_i =\frac{\pi}{6}\rho_i \sigma_i^3$ and the total packing fraction  $\eta =\frac{\pi}{6}\rho \sigma_b^3 (x_b  + x_s q^3)=\eta_b+\eta_s$. Then, from Eq.\ (\ref{Eq:OZ1}) with $\Nc=2$ one can
get the following results,
\begin{subequations}
\label{cq11-cq22}
\begin{equation}
\label{cq11}
\tilde{c}_{ss}({k})={{\tilde{h}_{ss}({k}) +\trho_b\left[\tilde{h}_{ss}({k}) \tilde{h}_{bb}({k})- \tilde{h}^2_{sb}({k})\right]} \over {\DD({k})}},
\end{equation}
\begin{equation}
\label{cq12}
\tilde{c}_{sb}({k})={{\tilde{h}_{sb}({k})} \over {\DD({k})}},
\end{equation}
\begin{equation}
\label{cq22}
\tilde{c}_{bb}({k})={{\tilde{h}_{bb}({k}) +\trho_s\left[\tilde{h}_{ss}({k}) \tilde{h}_{bb}({k})- \tilde{h}^2_{sb}({k})\right]} \over {\DD({k})}},
\end{equation}
\end{subequations}
where
\begin{eqnarray}
\DD({k})&=&1 + \rho_s \tilde{h}_{ss}({k}) + \rho_b \tilde{h}_{bb}({k})  \nonumber\\
&&+ \rho_s \rho_b \left[\tilde{h}_{ss}({k}) \tilde{h}_{bb}({k})-\tilde{h}^2_{sb}({k})\right].
\end{eqnarray}

Therefore, provided one can have accurate approximations of the Fourier transforms $\tilde{h}_{ij}({k})$ of the total correlation functions,  it is immediate to also obtain accurate approximations to the Fourier transforms $\tilde{c}_{ij}({k})$ from Eqs.\ (\ref{cq11-cq22}).  Finally, the DCFs $c_{ij}(r)$  are readily computed by taking inverse Fourier transforms:
\begin{equation}
c_{ij}({r})={1 \over {2 \pi^2}} \int_{0}^{\infty} d\tk\,\tk ^2 \tilde{c}_{ij}({k}) {{\sin{(\tk \tr)}} \over {\tk \tr}}.
\end{equation}

The large-$k$ behavior of $\tilde{c}_{ij}(k)$ has the structure \cite{PBYSH20}
\beq
\label{cijkn}
\tilde{c}_{ij}(k)\to\sum_{n=2}^\infty k^{-n}\tilde{c}_{ij}^{(n)}(k),
\eeq
where $\tilde{c}_{ij}^{(n)}(k)$ is a bound function expressed as a combination of sine and cosine functions of $\sigma_sk$, $\sigma_bk$, and $\sigma_{sb}k$. Thus, at a practical level, it is useful to introduce an arbitrarily large  cutoff wave number $Q$ and decompose $c_{ij}({r})$ into two contributions:
\begin{equation}
\label{cijr}
 c_{ij}({r})= c^{\num}_{ij}({r}) + c^{\tail}_{ij}({r}),
\end{equation}
with
\begin{subequations}
\label{Eq:crN-crA}
 \begin{equation}
\label{Eq:crN}
c^{\num}_{ij}({r})={1 \over {2 \pi^2}} \int_{0}^{Q} d\tk\,\tk ^2 \tilde{c}_{ij}({k}) {{\sin{(\tk \tr)}} \over {\tk \tr}}  ,
\end{equation}
\begin{equation}
\label{Eq:crA}
c^{\tail}_{ij}({r})={1 \over {2 \pi^2}} \sum_{n=2}^{n_{\max}}\int_{Q}^{\infty}d\tk\,  k^{-(n-2)}\tilde{c}_{ij}^{(n)}(k) {{\sin{(\tk \tr)}} \over {\tk \tr}}  ,
\end{equation}
\end{subequations}
where $n_{\max}$ is a conveniently chosen integer.
Assuming that $\tilde{c}_{ij}({k})$ is analytically known, the  contribution \eqref{Eq:crN} can be obtained numerically, whereas  the  contribution \eqref{Eq:crA} can be evaluated analytically term by term.

\subsection{WM scheme}
\label{WM}
The WM scheme \cite{PBYSH20,PYSHB21} allows one to obtain analytic forms for $\tilde{c}_{ij}(k)$. The method relies on accurate MD simulation data (here obtained via the DynamO program \cite{BSL11}) for the total correlation functions in combination with  their pole structure representation  and the OZ equation.

The MD simulation data $h_{ij}^{\text{MD}}({r})$ are fitted to the  semiempirical approximation
\begin{equation}
\label{Eq:hrWM}
h^{WM}_{ij}({r}) =
\begin{cases}
-1,  \qquad ~~~~~~~~~~~~~~~~~~~~~~~~~~~~~~ 0 < \tr < \sigma_{ij}, \\
\sum\limits_{n=1}^{W} b_{ij}^{(n)} \tr^{n-1},\qquad ~~~~~~~~~~~~~~ \sigma_{ij} < \tr \leq \tr_{ij}^{{m}}, \\
\sum\limits_{n=1}^{M} \frac{\tA_{ij}^{(n)}}{\tr} e^{-\talpha_{n} \tr} \sin\left(\tomega_{n} \tr + \tdelta_{ij}^{(n)}\right) ,\quad  r\geq\tr_{ij}^{{m}}.
\end{cases}
\end{equation}
where $\tr_{ij}^{{m}}$ is chosen as the position of the first minimum of $h_{ij}^{WM}({r})$, and $\{ b_{ij}^{(n)}; n=1,\ldots, W \}$ and $\{ \tA_{ij}^{(n)}, \talpha_{n}, \tomega_{n}, \tdelta_{ij}^{(n)}; n=1,\ldots,M  \}$ are fitting parameters. Moreover, the
continuity of $h_{ij}^{WM}(r)$ and their first derivatives at $r=\tr_{ij}^{{m}}$, as well as  the Boubl\'ik-Grundke-Henderson-Lee-Levesque (BGHLL) contact values \cite{B70,GH72,LL73}, are enforced. Convenient choices for the parameters $W$ and $M$ are
$W=15$ and $M=10$. 
To get sufficient accuracy, the data for $h_{ij}^{MD} (r)$ were obtained from long simulations ($ \sim 10^9$ total collisions in production) and for a large number of particles (typically $16384$--$48668$ particles, depending on density and the size ratio). For further details on the MD simulations and the WM scheme, the reader is referred to Refs.\ \cite{PBYSH20,PYSHB21}.

From the parametrization \eqref{Eq:hrWM}, it is possible to obtain the Fourier transforms $\tilde{h}_{ij}^{WM}(k)$ analytically \cite{PBYSH20}. Next, the OZ relations \eqref{cq11-cq22} yield analytic expressions for $\tilde{c}_{ij}^{WM}(k)$, from which the associated asymptotic functions $\tilde{c}_{ij}^{WM(n)}(k)$ follow [see Eq.\ \eqref{cijkn}]. Finally, the DCFs are obtained from Eqs.\ \eqref{cijr} and \eqref{Eq:crN-crA} with the choices $Q=200/\sigma_{ij}$ and $n_{\max}=6$. In summary,
\bal
h_{ij}^{\text{MD}}(r)&\stackrel{\text{Eq.\ \eqref{Eq:hrWM}}}{\longrightarrow}h_{ij}^{WM}(r)\stackrel{\text{Eq.\ \eqref{FT}}}{\longrightarrow} \tilde{h}_{ij}^{WM}(k)\nn
&\stackrel{\text{Eqs.\ \eqref{cq11-cq22}}}{\longrightarrow}\tilde{c}_{ij}^{WM}(k)\stackrel{\text{Eqs.\  \eqref{cijr}--\eqref{Eq:crN-crA}}}{\longrightarrow}c_{ij}^{WM}(r).
\eal

\subsection{Rational-function approximation}
\label{secRFA}
We shall now sketch the RFA approach to obtain the structural properties of additive HS mixtures. The detailed description  may be found elsewhere \cite{YSH98,HYS08,S16,SYH20,note_19_07}. First, we introduce the Laplace transforms of $r g_{ij}(r)$:
\beq
\label{3.1}
G_{ij}(z)=\int_0^\infty d r\, e^{-zr}r g_{ij}(r).
\eeq
The Fourier transform $\widetilde{h}_{ij}(k)$ is related to $G_{ij}(z)$ by
\beq
\widetilde{h}_{ij}(k)=-2\pi \left.\frac{G_{ij}(z)-G_{ij}(-z)}{z}
\right|_{z=\imath k},
\label{1.7}
\end{equation}
$\imath$ being the imaginary unit.
Next, we propose the following form for $G_{ij}(z)$:
\beq
\label{3.6}
G_{ij}^{\text{RFA}}(z)=\frac{e^{-\sigma_{ij} z}}{2\pi z^2}
\left({\sf L}(z)\cdot \left[(1+\xi z)\openonebis-{ \sf
A}(z)\right]^{-1}\right)_{ij},
\eeq
\begin{subequations}
where $\openonebis$ is the unit matrix, $\xi$ is a parameter to be fixed, and
\beq
\label{3.7}
L_{ij}(z)=L_{ij}^\zero+L_{ij}^\one
z+L_{ij}^\two z^2,
\eeq
\bal
\label{3.8}
A_{ij}(z)=&\rho_i\left[\varphi_2(\sigma_i z)\sigma_i^3 L_{ij}^\zero
+\varphi_1(\sigma_i z)\sigma_{i}^2 L_{ij}^\one\right.\nn
&\left. +\varphi_0(\sigma_{i}
z)\sigma_{i} L_{ij}^\two\right],
\eal
\end{subequations}
the functions $\varphi_n(x)$ being defined by
\beq
\label{2.9}
\varphi_n(x)\equiv x^{-(n+1)}\left(\sum_{m=0}^n \frac{(-x)^m}{m!}-
e^{-x}\right).
\eeq
Then, by imposing certain
consistency conditions, the elements
of the matrices $\mathsf{L}^\zero$, $\mathsf{L}^\one$, $\mathsf{L}^\two$ are expressed as linear functions of $\xi$. In particular, $L_{ij}^\two=2\pi\xi\sigma_{ij} g_{ij}^c$, where  $g_{ij}^c\equiv g(\sigma_{ij}^+)$ are the contact values of the radial distribution functions.

The special choice $\xi=0$ gives the PY solution  \cite{L64,BH76}. On the other hand, by an appropriate determination of $\xi\neq 0$ as the physical root of a polynomial equation, the RFA can be made thermodynamically consistent  and, additionally, allows one to freely choose the contact values $g_{ij}^c$, a convenient choice being the BGHLL expression \cite{B70,GH72,LL73}.

Once $G_{ij}^{\text{RFA}}(z)$ is analytically known, $\tilde{h}_{ij}^{\text{RFA}}(k)$ can be obtained from application of the exact relationship \eqref{1.7}. From here, the procedure is similar to the WM case:
analytic expressions for $\tilde{c}_{ij}^{\text{RFA}}(k)$  are obtained from Eqs.\ \eqref{cq11-cq22}, from which one gets the asymptotic functions $\tilde{c}_{ij}^{\text{RFA}(n)}(k)$; then, the RFA DCFs are numerically obtained by application of Eqs.\ \eqref{cijr} and \eqref{Eq:crN-crA} again with the choices $Q=200/\sigma_{ij}$ and $n_{\max}=6$. Thus,
\bal
G_{ij}^{\text{RFA}}(z)&\stackrel{\text{Eq.\ \eqref{1.7}}}{\longrightarrow} \tilde{h}_{ij}^{\text{RFA}}(k)\nn
&\stackrel{\text{Eqs.\ \eqref{cq11-cq22}}}{\longrightarrow}\tilde{c}_{ij}^{\text{RFA}}(k)\stackrel{\text{Eqs.\  \eqref{cijr}--\eqref{Eq:crN-crA}}}{\longrightarrow}c_{ij}^{\text{RFA}}(r).
\eal

\section{Results}
\label{sec3}
We now present the results of the comparison between the DCFs predicted by the RFA with those obtained via the WM scheme from our MD simulations.

\begin{figure}
\includegraphics[width=0.8\columnwidth]{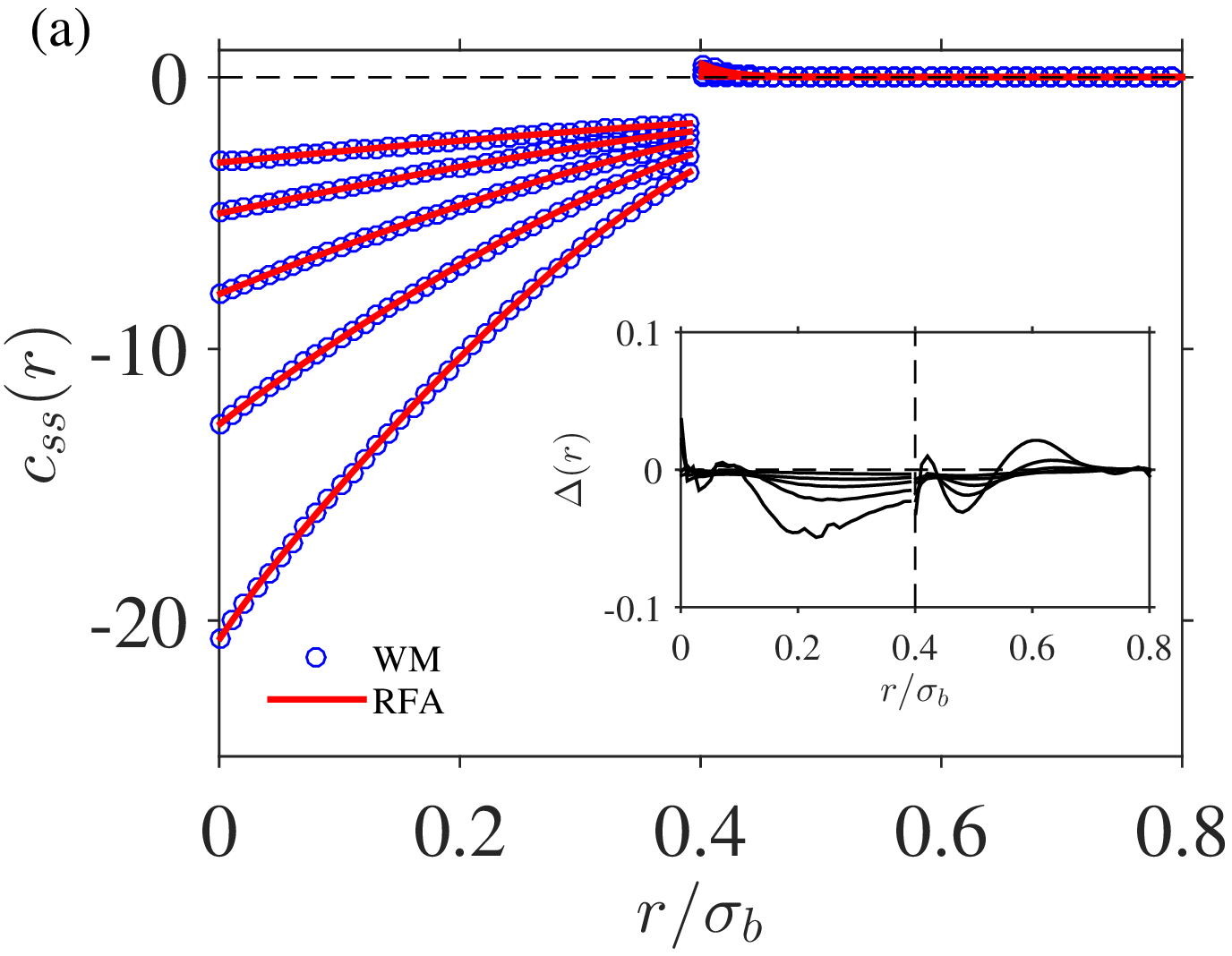}\\
\includegraphics[width=0.83\columnwidth]{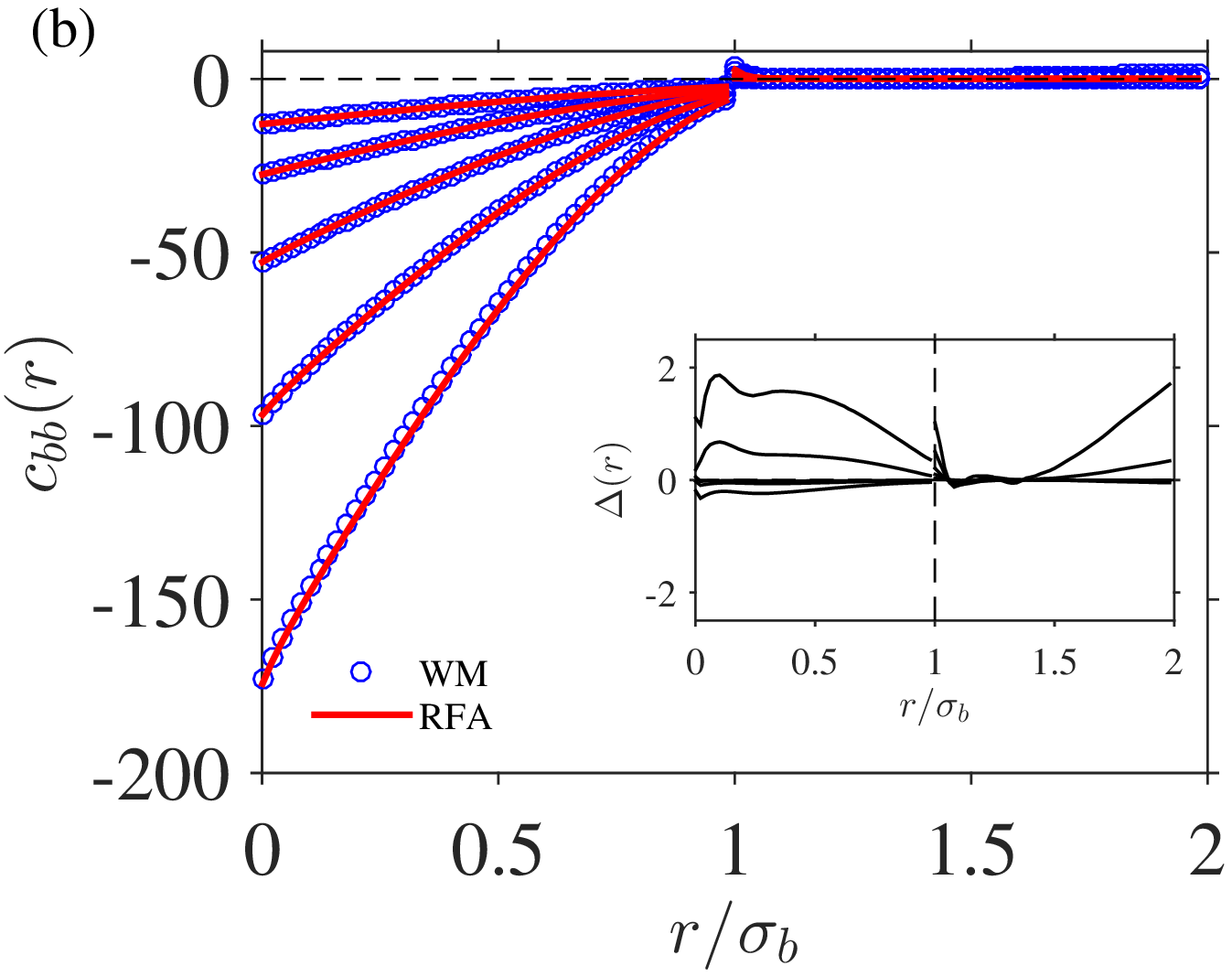}\\
\includegraphics[width=0.8\columnwidth]{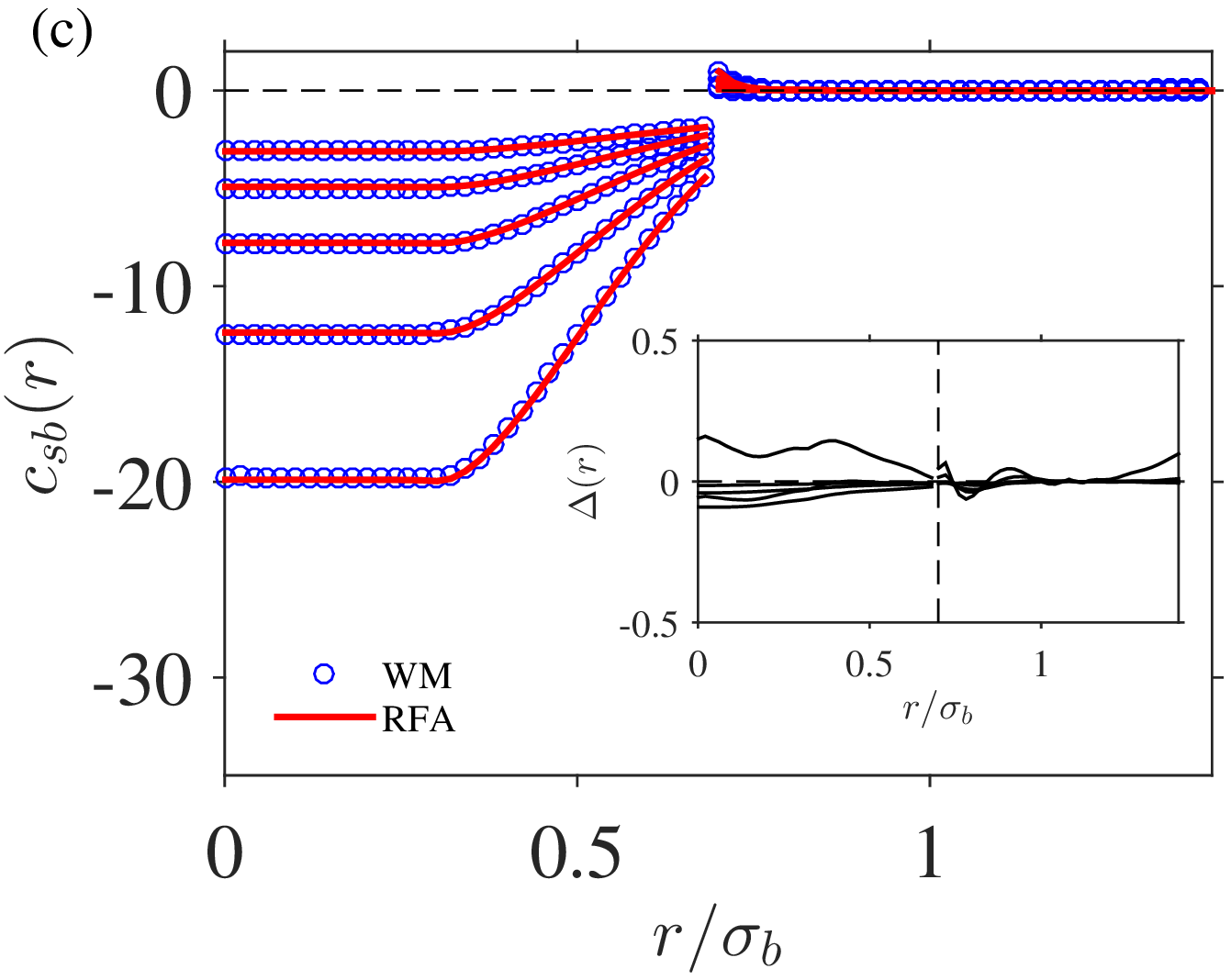}
\caption{Plot of (a) $c_{ss}(r)$, (b) $c_{bb}(r)$, and (c) $c_{sb}(r)$ for a size ratio $q=0.4$, a partial packing fraction $\eta_b=0.2$, and (from top to bottom in each panel) $\eta_s=0.05$, $0.1$, $0.15$, $0.2$, and $0.25$. The blue circles are the WM results and the red thick lines correspond to the RFA values. The insets show the differences $\Delta(r) =c_{ij}^{WM} (r)-c_{ij}^{\text{RFA}} (r)$, which tend to increase with increasing $\eta_s$. }
\label{fig1}
\end{figure}

Figure \ref{fig1} shows the comparison for the rather disparate BHS mixture $q=0.4$ at fixed $\eta_b=0.2$ and several values of $\eta_s$. For all three  DCFs,  we observe very good agreement between the WM method and the RFA, once more confirming our previous findings \cite{PBYSH20,PYSHB21}.  In fact, the relative differences for the DCFs are small (usually smaller than $2$--$4\%$ in the core region, depending on density), even for the densest cases. In contrast, the differences with the PY theory are generally quite significant (not shown).

We also observe that the like-like functions $c_{ss} (r<\sigma_s)$ and $c_{bb} (r<\sigma_b)$ inside the core are monotonic and can be well represented by a low-order polynomial (usually, fourth or sixth degree is sufficient). Also, their limiting values $c_{ss} (r=0)$, $c_{bb} (r=0)$, $c_{ss} (r=\sigma_s^-)$, and $c_{bb} (r=\sigma_b^-)$ can be determined fairly accurately. On the other hand, while hardly apparent in Fig.\ \ref{fig1}(c), both approaches (the WM method and the RFA) indicate that the cross function $c_{sb} (r<\sigma_{sb})$ is not monotonic inside the core, what requires a separate more detailed analysis to be carried out below. Outside the core, the functions $c_{ss} (r>\sigma_s)$, $c_{bb} (r>\sigma_b)$, and $c_{sb} (r>\sigma_{sb})$ are oscillatory decaying for the WM method, monotonically decaying for the RFA,  and vanishing for the PY theory \cite{PBYSH20}. As discussed in Ref.\ \cite{PBYSH20},   the part of the DCFs outside the core, namely,  $c_{ij} (r>\sigma_{ij})$, is quite important for the asymptotic behavior of the total correlation functions, and  cannot be omitted as in the PY theory.

\begin{figure}
\includegraphics[width=0.8\columnwidth]{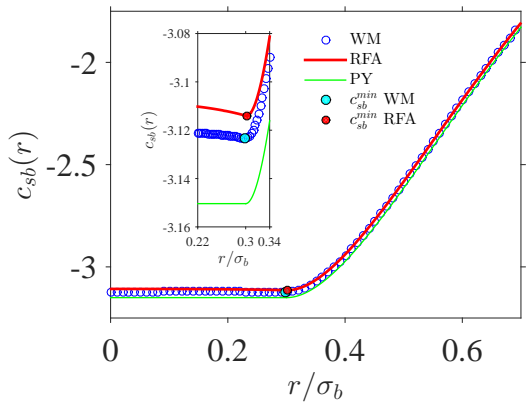}
\caption{Plot of  $c_{sb}(r)$ inside the core ($r<\sigma_{sb}$) for a size ratio $q=0.4$ and partial packing fractions $\eta_b=0.2$ and $\eta_s=0.05$. The blue circles are the WM results, the red thick lines correspond to the RFA values, and the green thin lines represent the PY values. The inset shows details of the curves in the range $0.22\leq r/\sigma_b\leq 0.34$. The cyan and red solid circles indicate the position of the minimum for WM and RFA, respectively.}
\label{fig2}
\end{figure}

\begin{figure*}
	\includegraphics[width=.63\columnwidth]{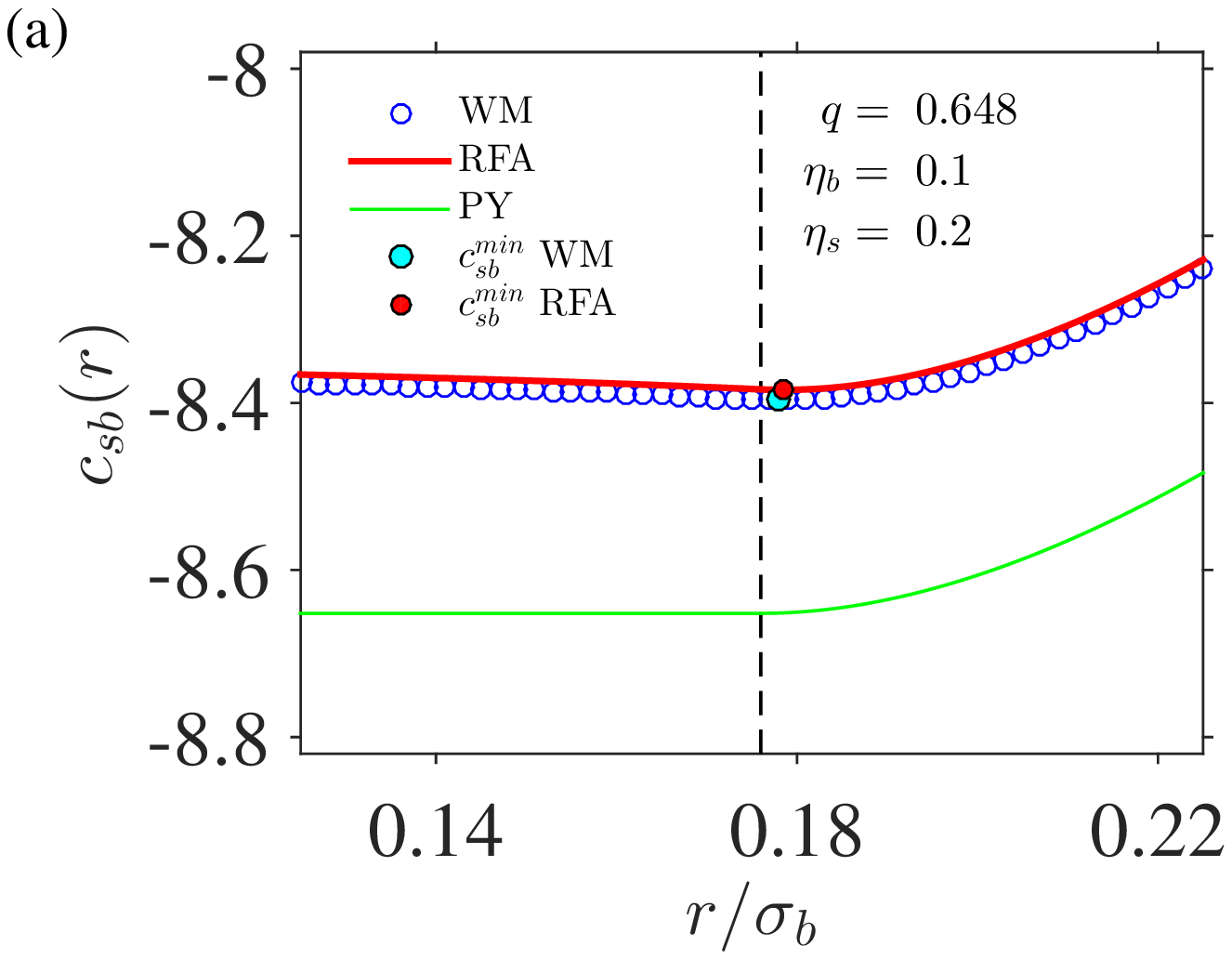}
	\includegraphics[width=.655\columnwidth]{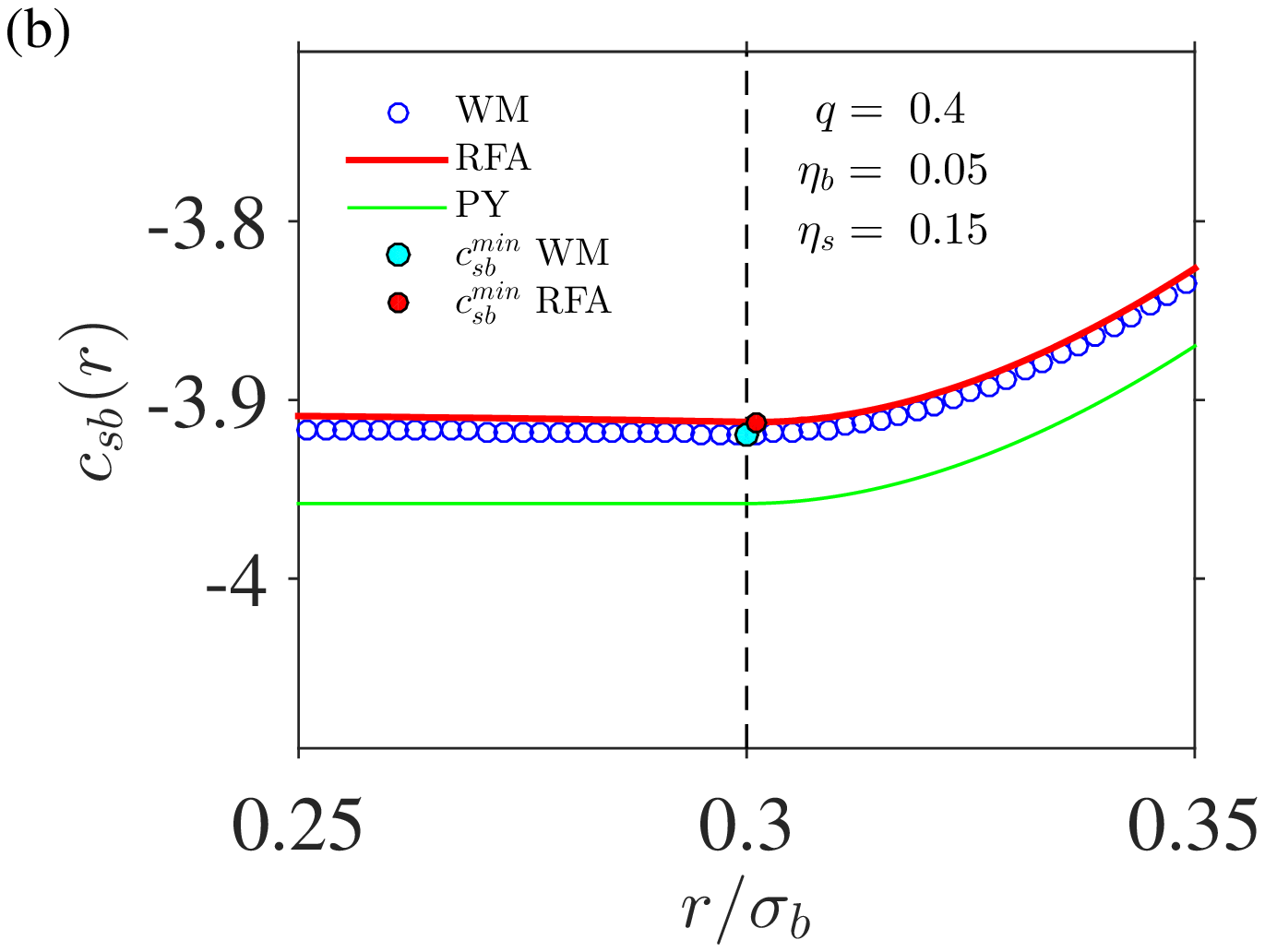}
	\includegraphics[width=.65\columnwidth]{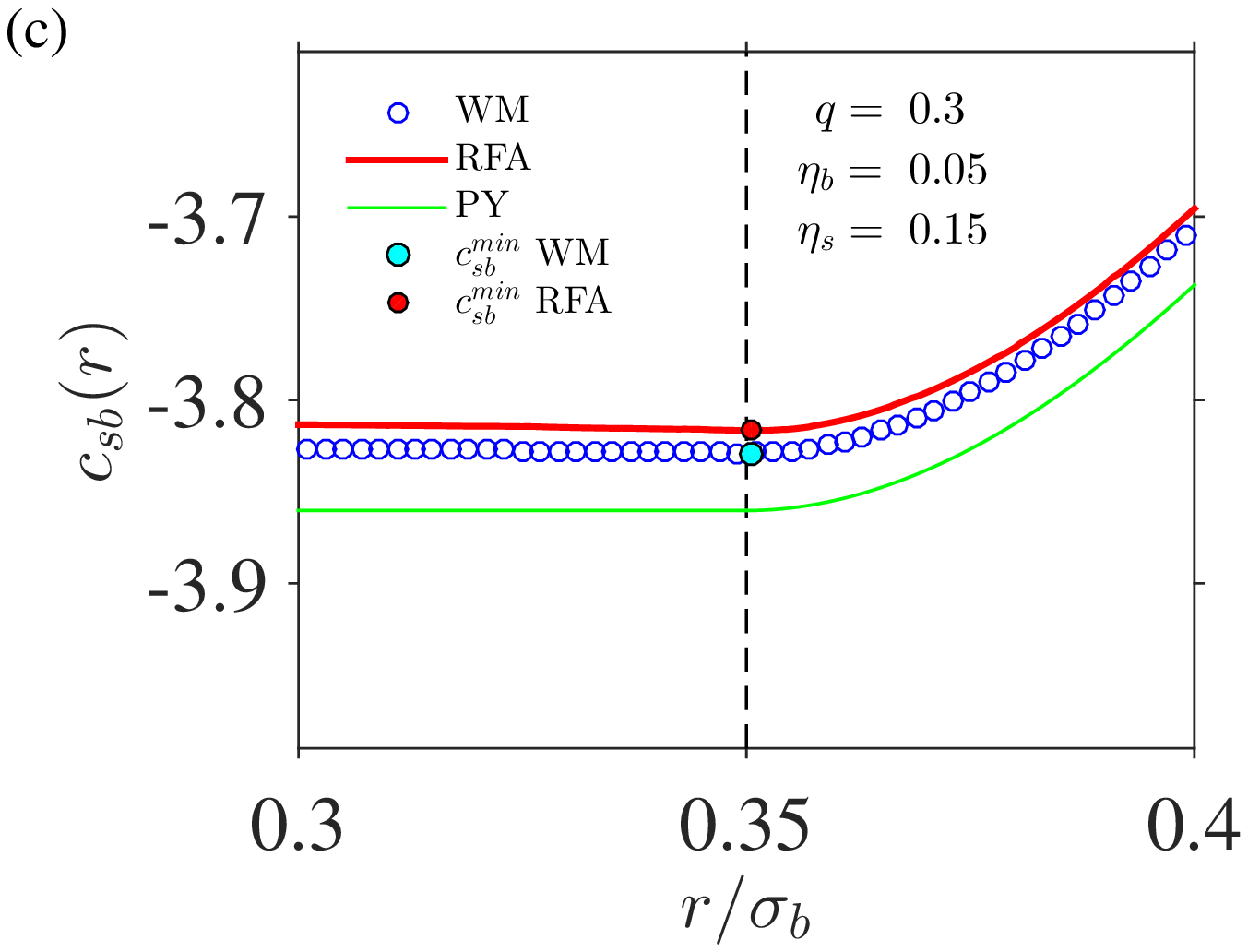}\\
	\includegraphics[width=.63\columnwidth]{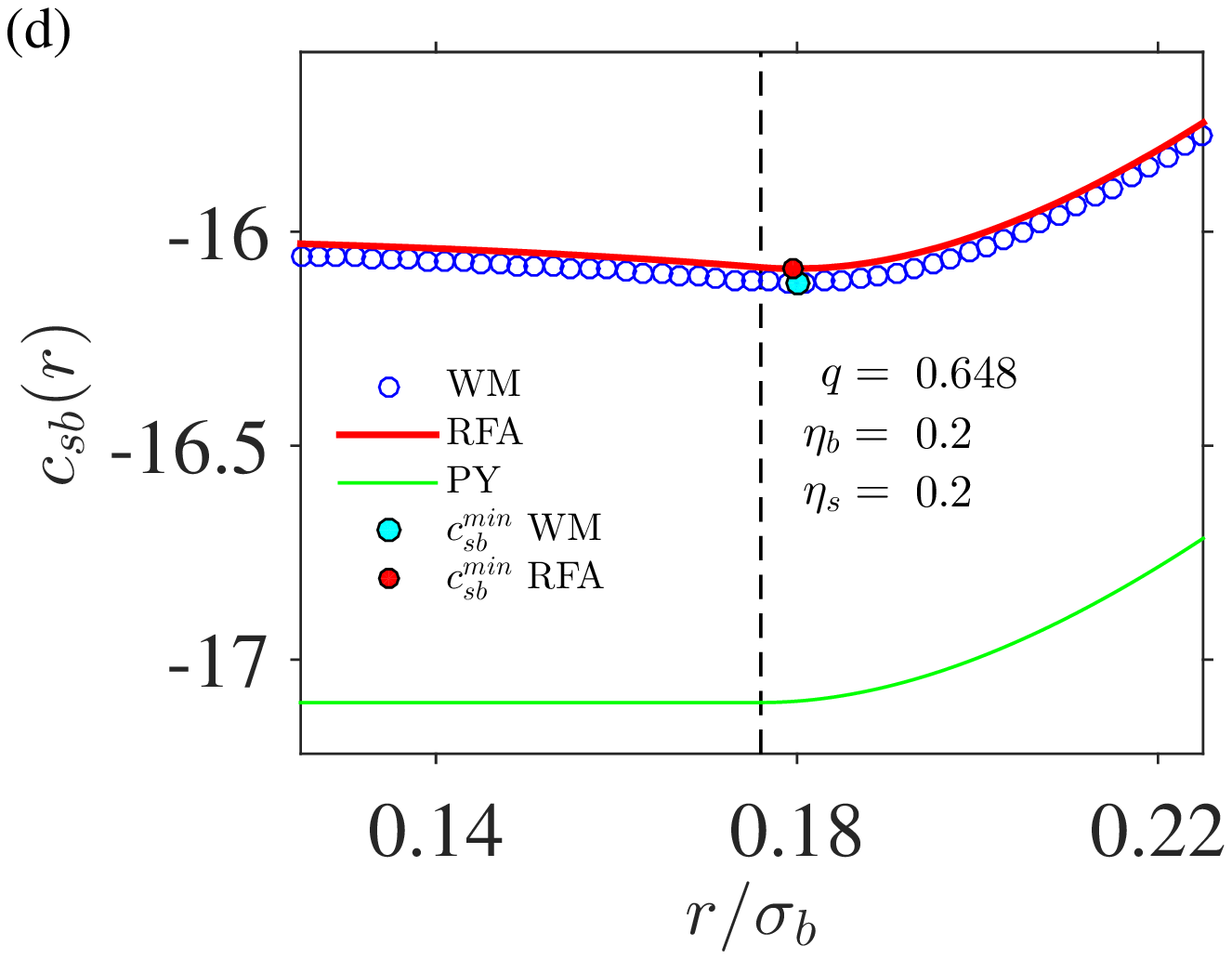}
	\includegraphics[width=.655\columnwidth]{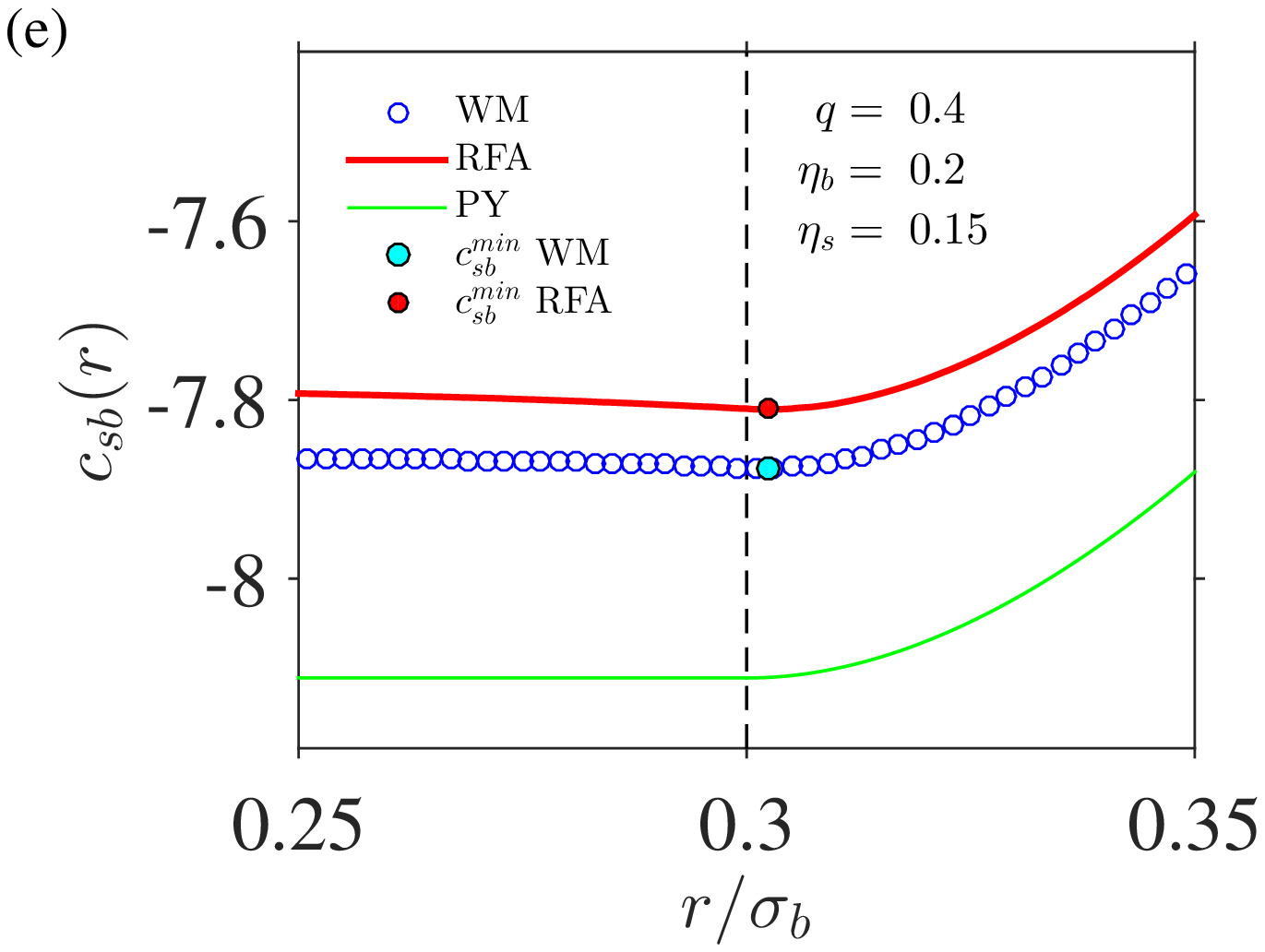}
	\includegraphics[width=.65\columnwidth]{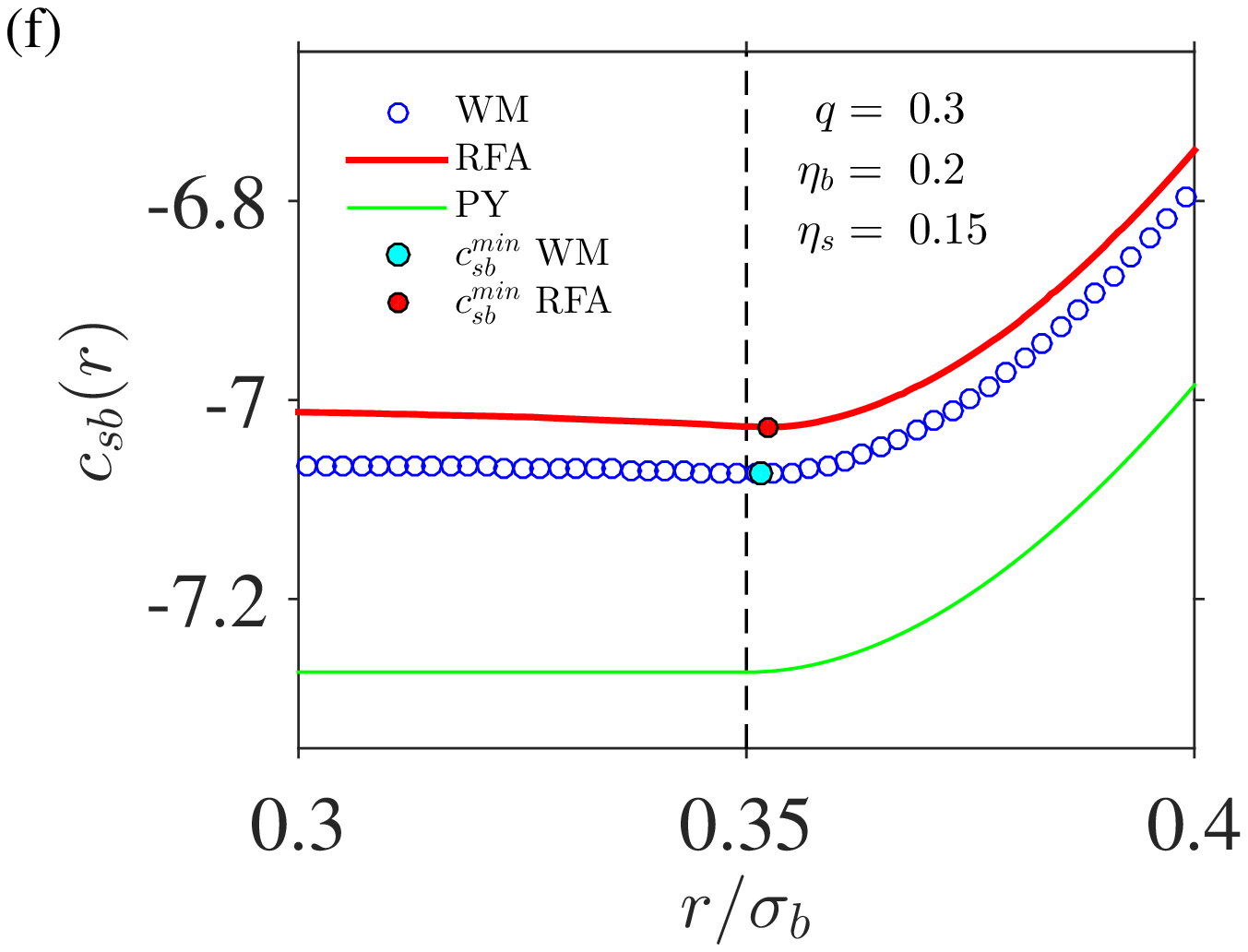}
	\caption{Plot of  $c_{sb}(r)$ around $r=\lambda_{sb}=\frac{1}{2}(\sigma_b-\sigma_s)$ for (a) $q=0.648$, $\eta_b=0.1$, $\eta_s=0.2$,  (b) $q=0.4$, $\eta_b=0.05$, $\eta_s=0.15$,  (c) $q=0.3$, $\eta_b=0.05$, $\eta_s=0.15$,  (d) $q=0.648$, $\eta_b=0.2$, $\eta_s=0.2$, (e) $q=0.4$, $\eta_b=0.2$, $\eta_s=0.15$, and (f) $q=0.3$, $\eta_b=0.2$, $\eta_s=0.15$. The blue circles are the WM results, the red thick lines correspond to the RFA values, and the green thin lines represent the PY values. In each panel, the cyan and red solid circles indicate the position of the minimum for WM and RFA, respectively, while the vertical dashed line signals the location of $\lambda_{sb}$. In each panel, the vertical axis range was selected as $\pm 5\%$ of the minimum value.}
	\label{fig2a}
\end{figure*}

\begin{figure*}
	\includegraphics[width=.65\columnwidth]{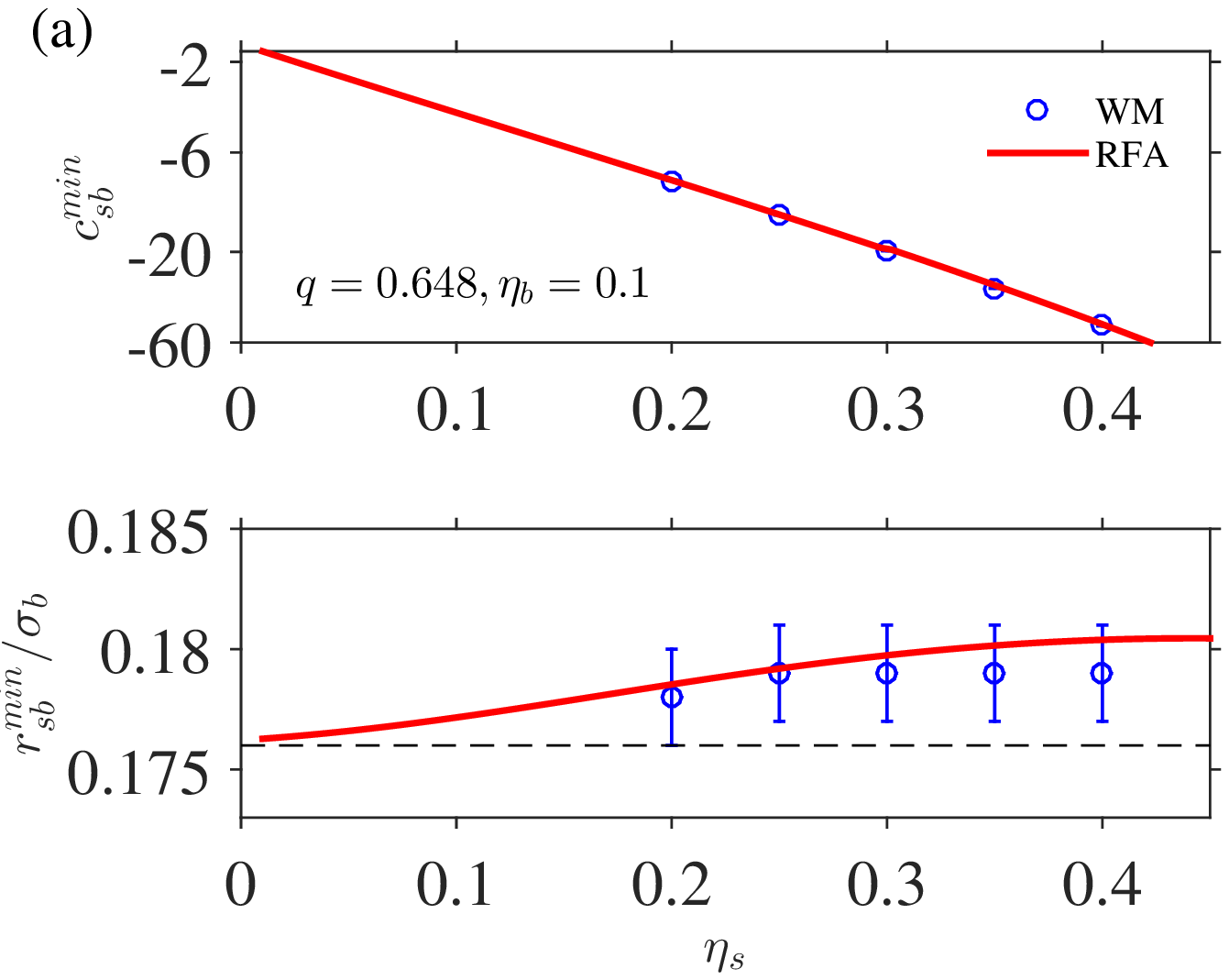}
	\includegraphics[width=.65\columnwidth]{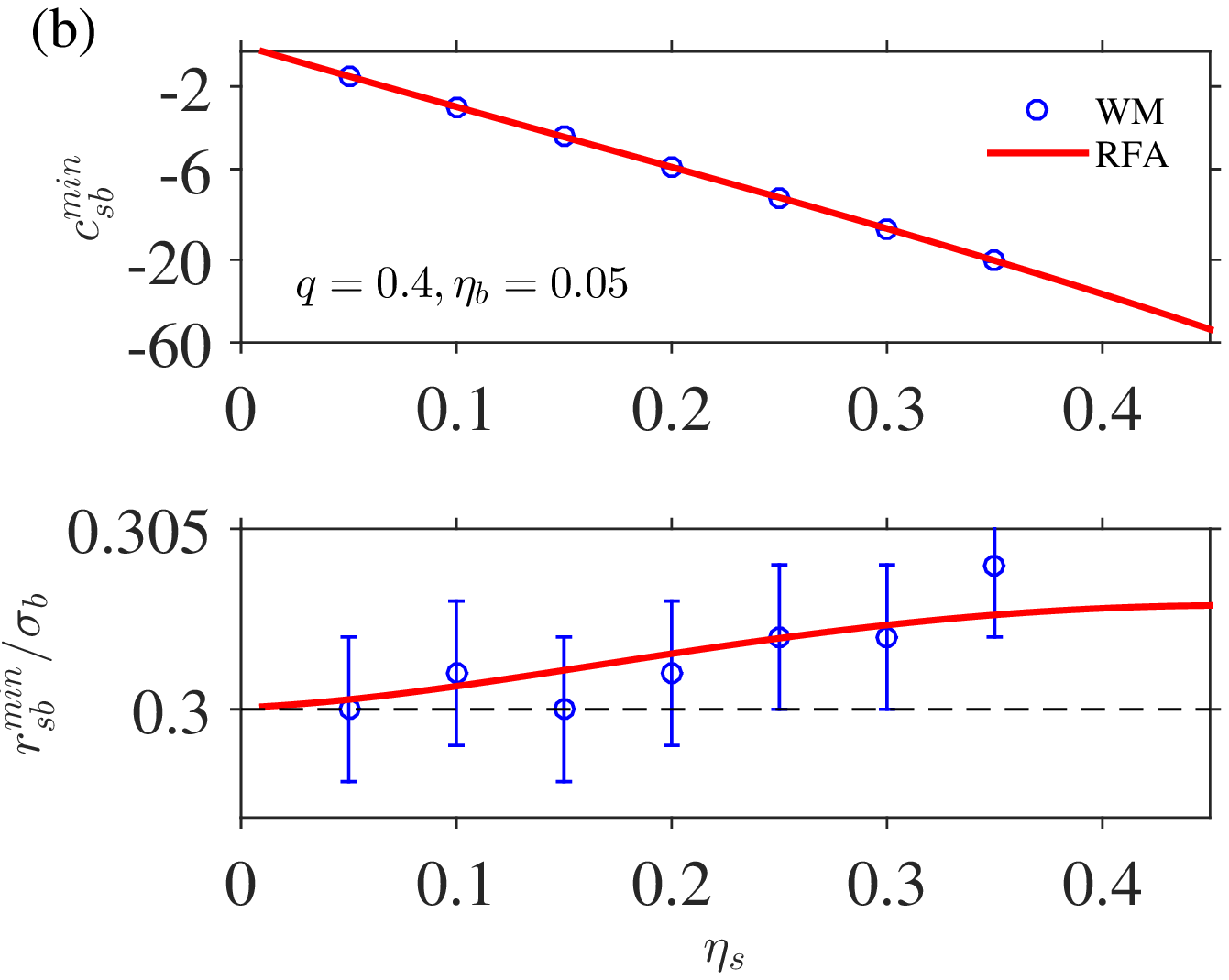}
	\includegraphics[width=.65\columnwidth]{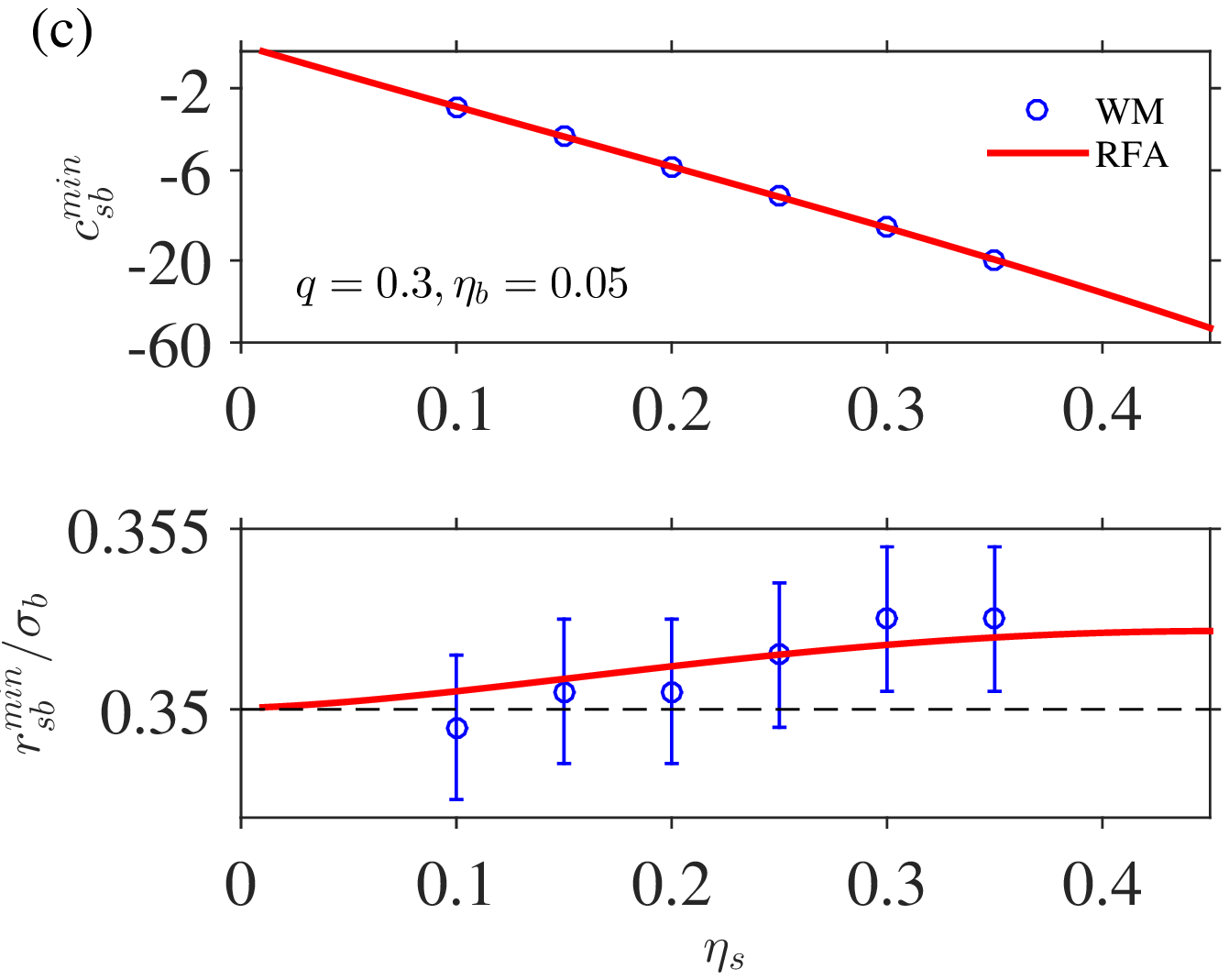}\\
	\includegraphics[width=.65\columnwidth]{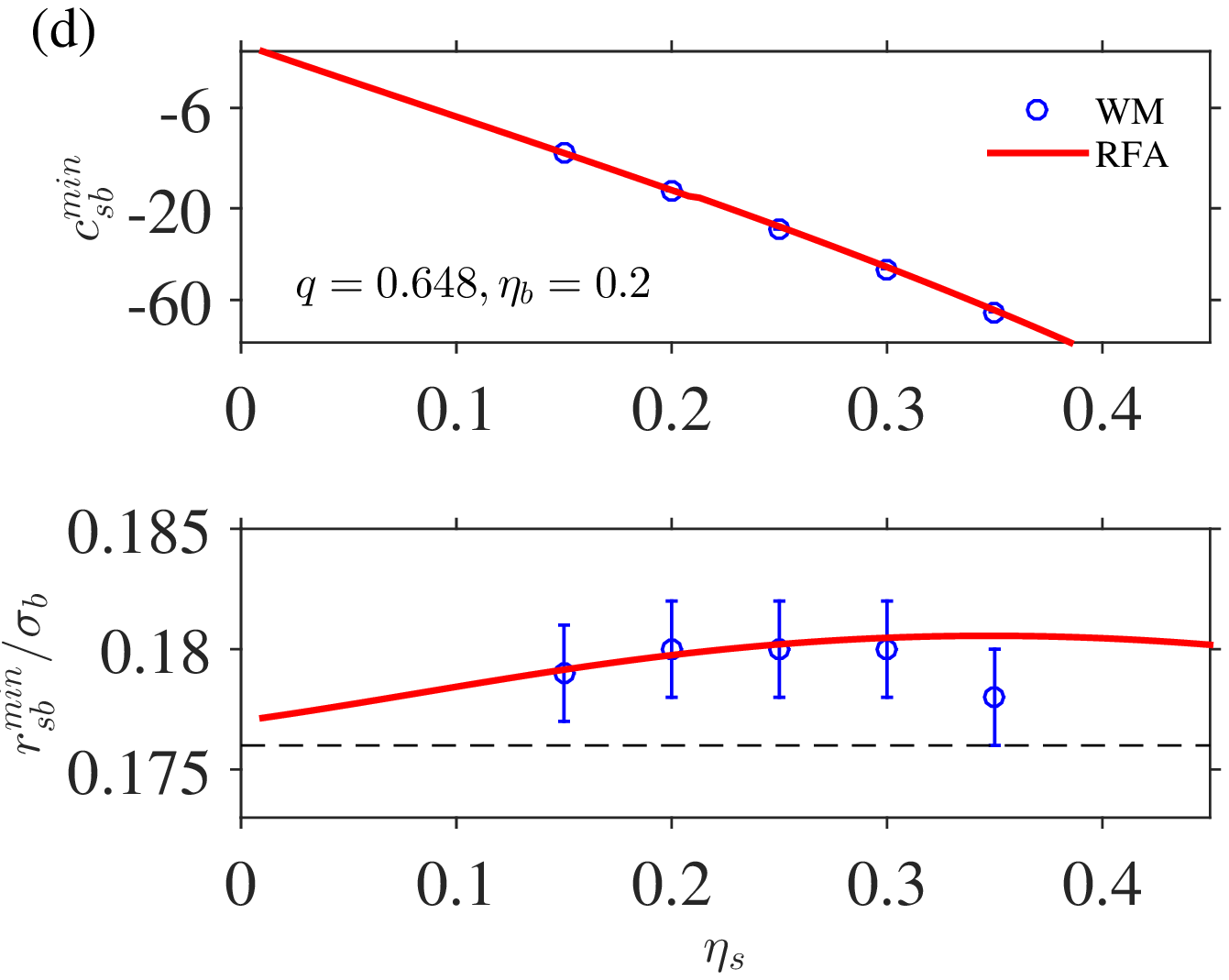}
	\includegraphics[width=.65\columnwidth]{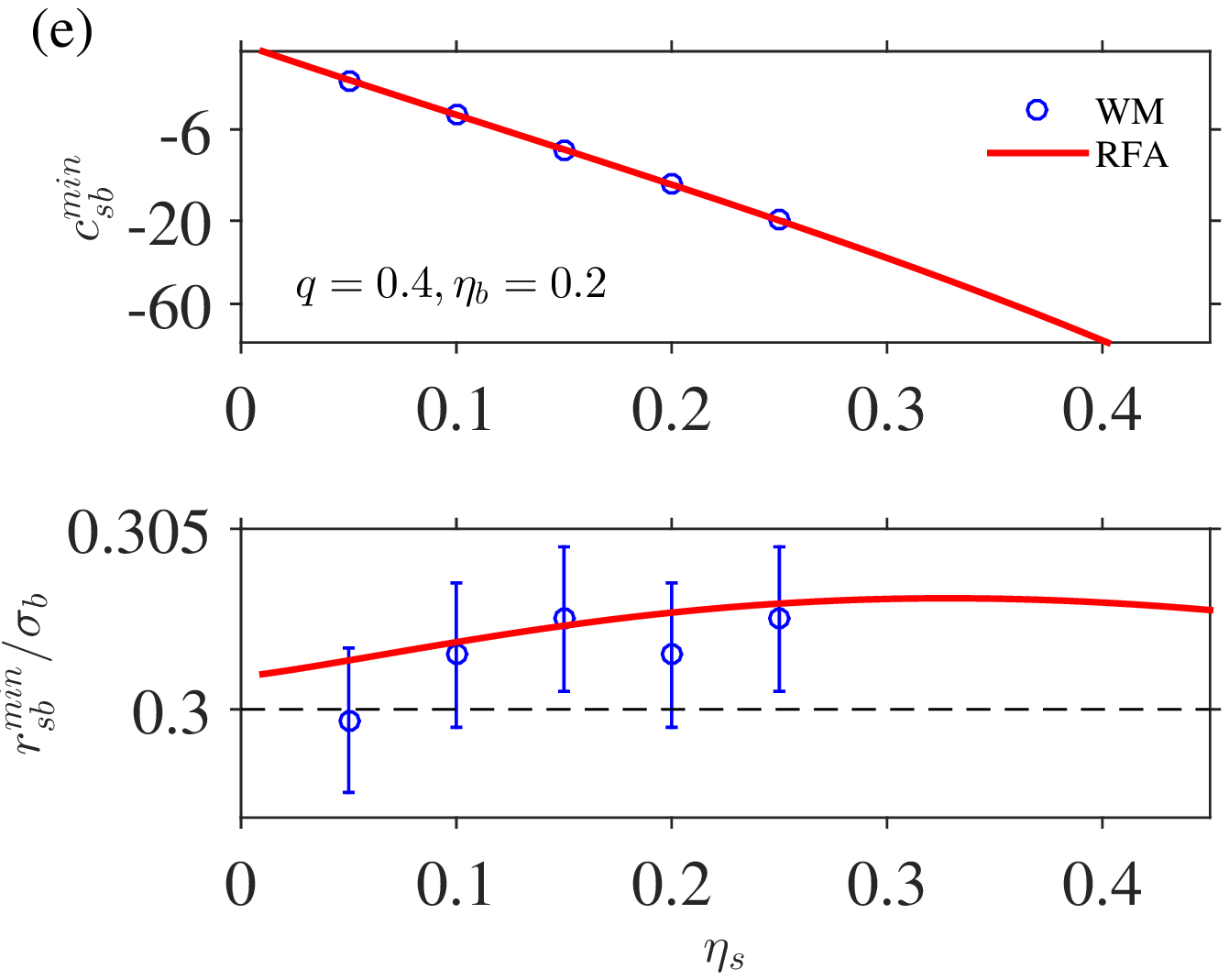}
	\includegraphics[width=.65\columnwidth]{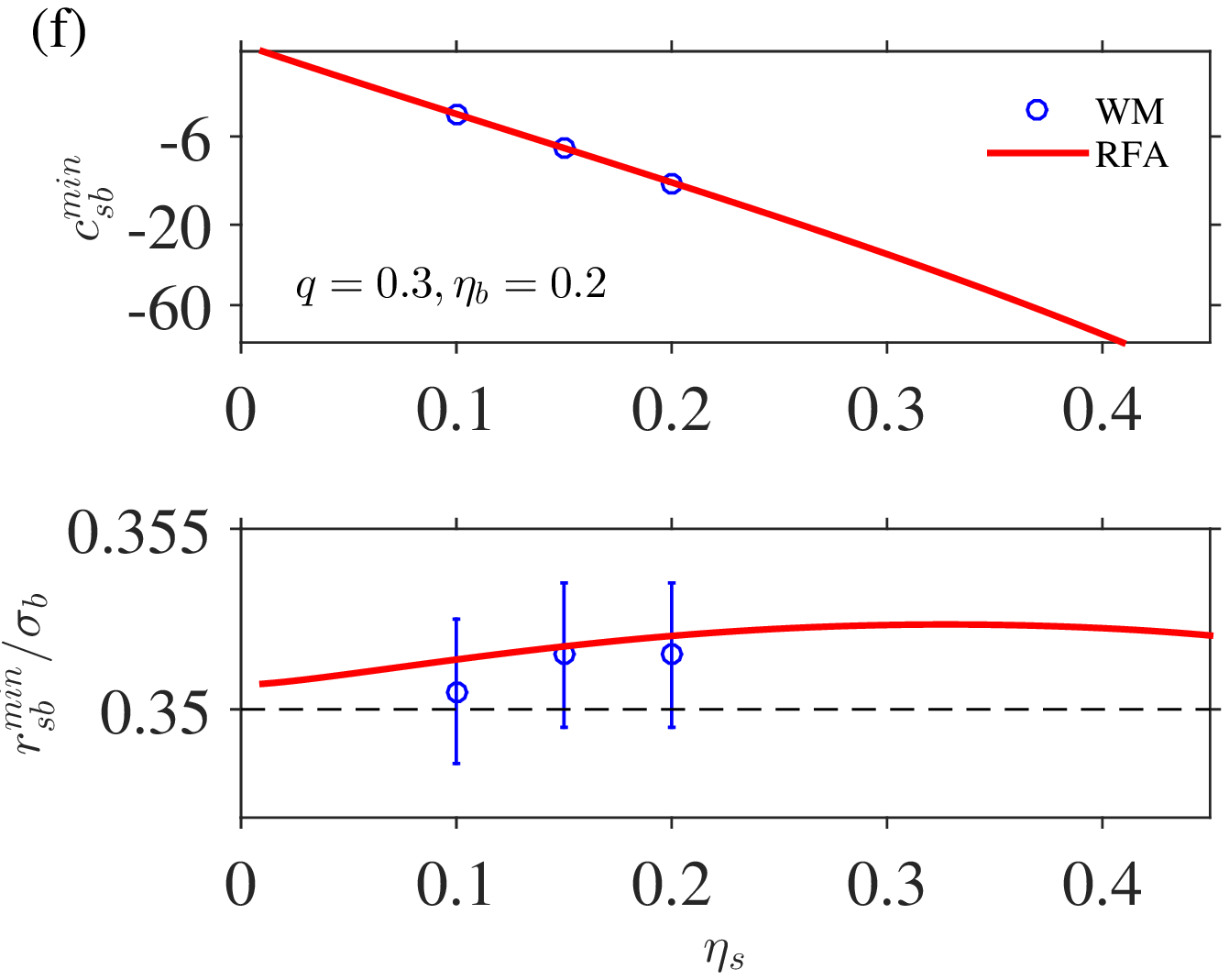}
	\caption{Dependence of $c_{sb}^{\min}$ (in logarithmic scale) and $r_{sb}^{\min}$ on $\eta_s$ for (a) $q=0.648$, $\eta_b=0.1$,  (b) $q=0.4$, $\eta_b=0.05$,  (c) $q=0.3$, $\eta_b=0.05$,  (d) $q=0.648$, $\eta_b=0.2$, (e) $q=0.4$, $\eta_b=0.2$, and (f) $q=0.3$, $\eta_b=0.2$. The blue circles are the WM results and the red thick lines correspond to the RFA values. In each plot of $r_{sb}^{\min}$, the horizontal dashed line signals the location of $\lambda_{sb}$.}
	\label{fig3}
\end{figure*}

\begin{figure*}
\includegraphics[width=.67\columnwidth]{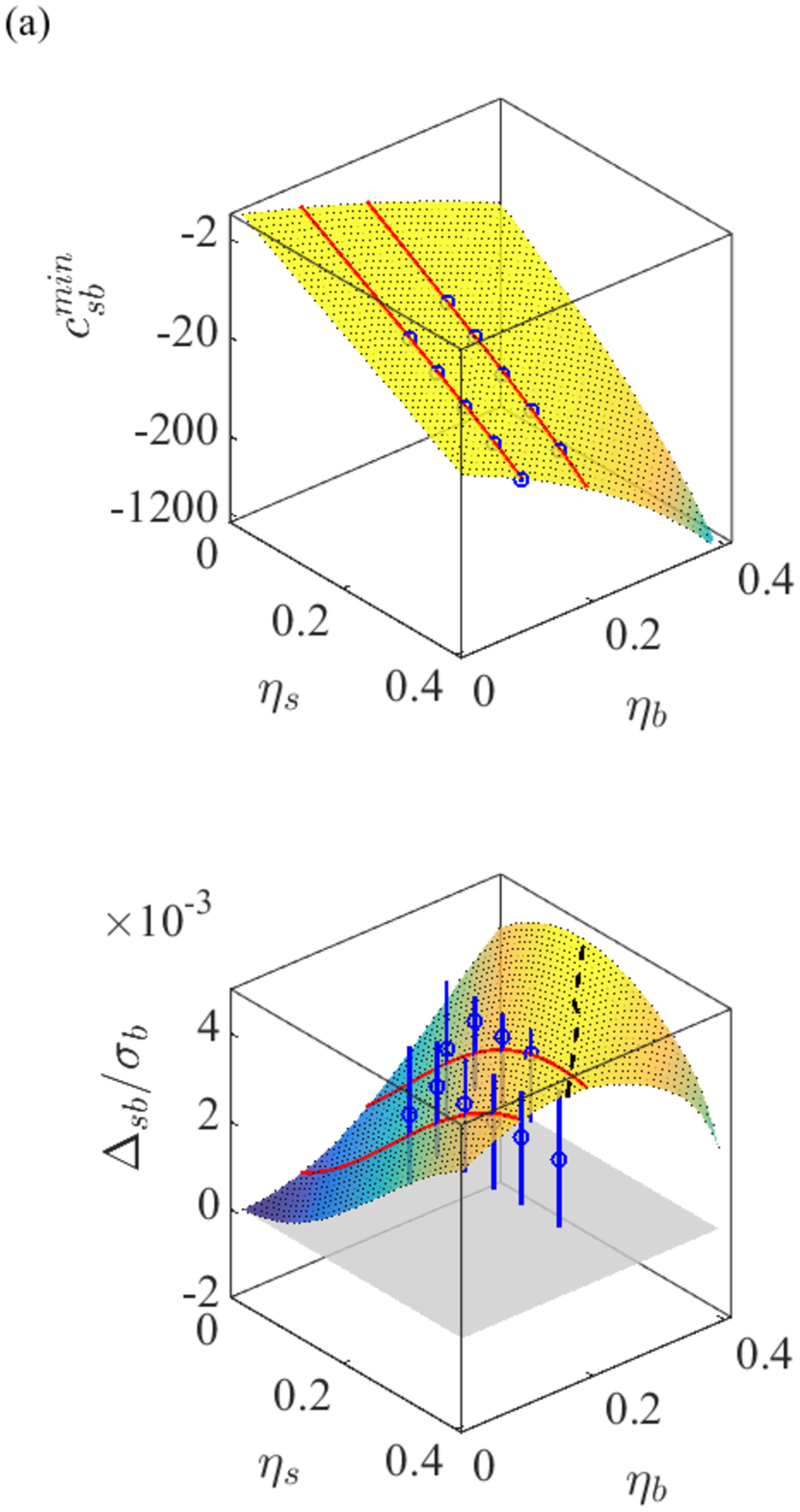}
\includegraphics[width=.67\columnwidth]{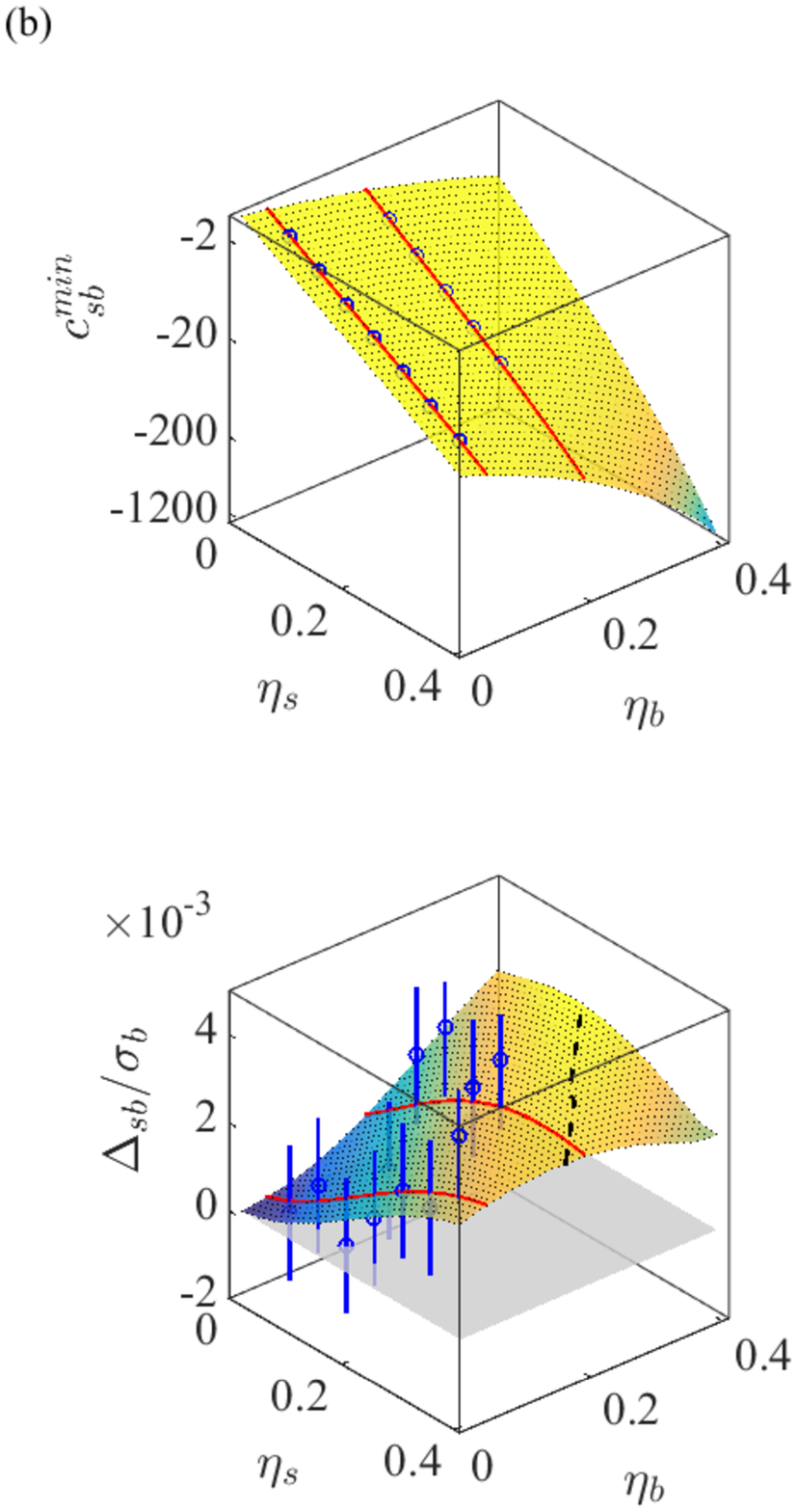}
\includegraphics[width=.67\columnwidth]{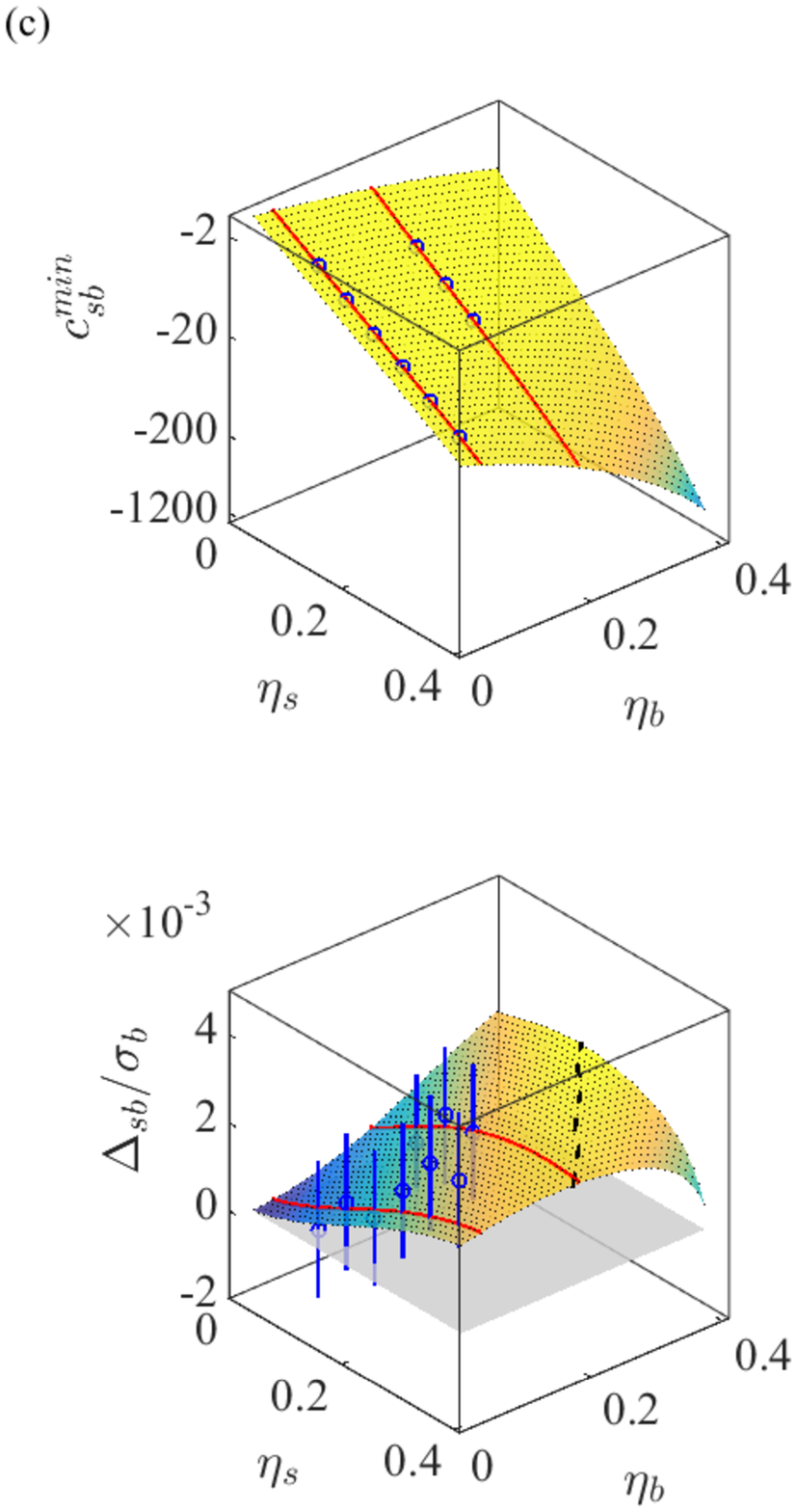}
\caption{3D plots, as predicted by the RFA, showing the density dependence of $c_{sb}^{\min}$ (in logarithmic scale) and the difference $\Delta_{sb}\equiv r_{sb}^{\min}-\lambda_{sb}$ for (a) $q=0.648$,  (b) $q=0.4$, and (c) $q=0.3$. The red solid lines represent the cases presented in Fig.\ \ref{fig3}, and the blue circles (with error bars) are the results from the WM scheme.  The black dashed lines show the maximum values of $\Delta_{sb}$ at fixed $\eta_b$. }
\label{fig4}
\end{figure*}

\begin{figure*}
\includegraphics[width=.65\columnwidth]{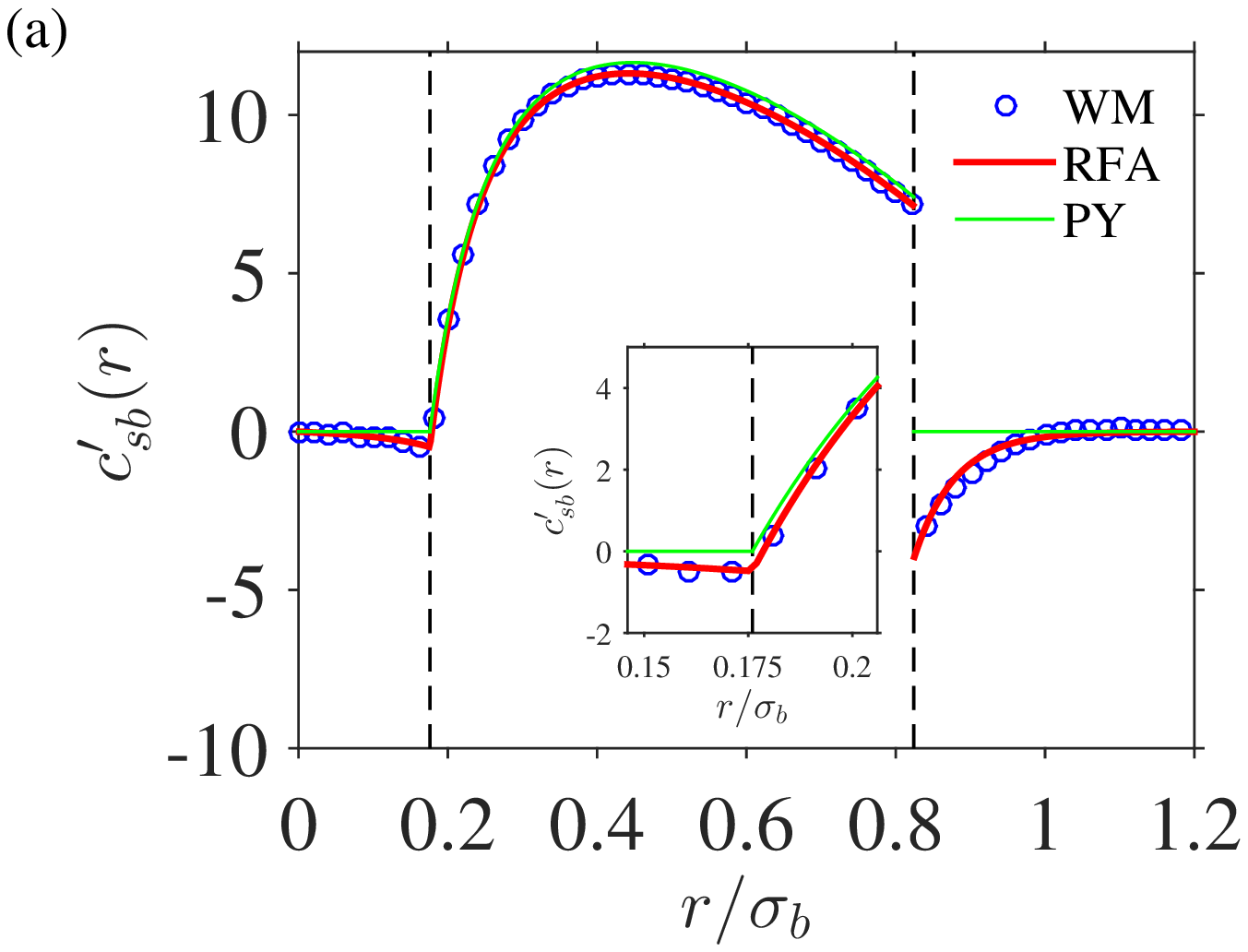}
\includegraphics[width=.63\columnwidth]{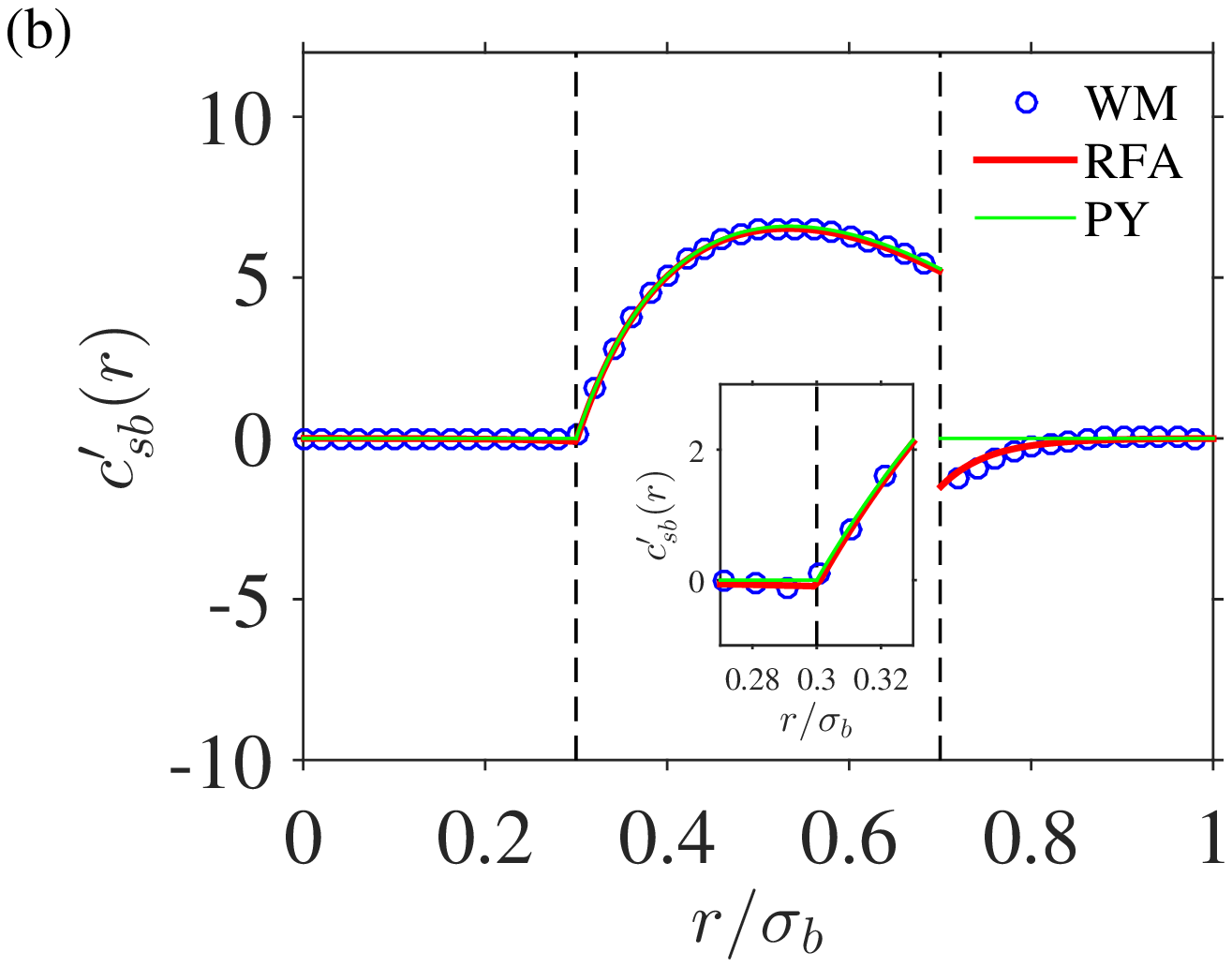}
\includegraphics[width=.65\columnwidth]{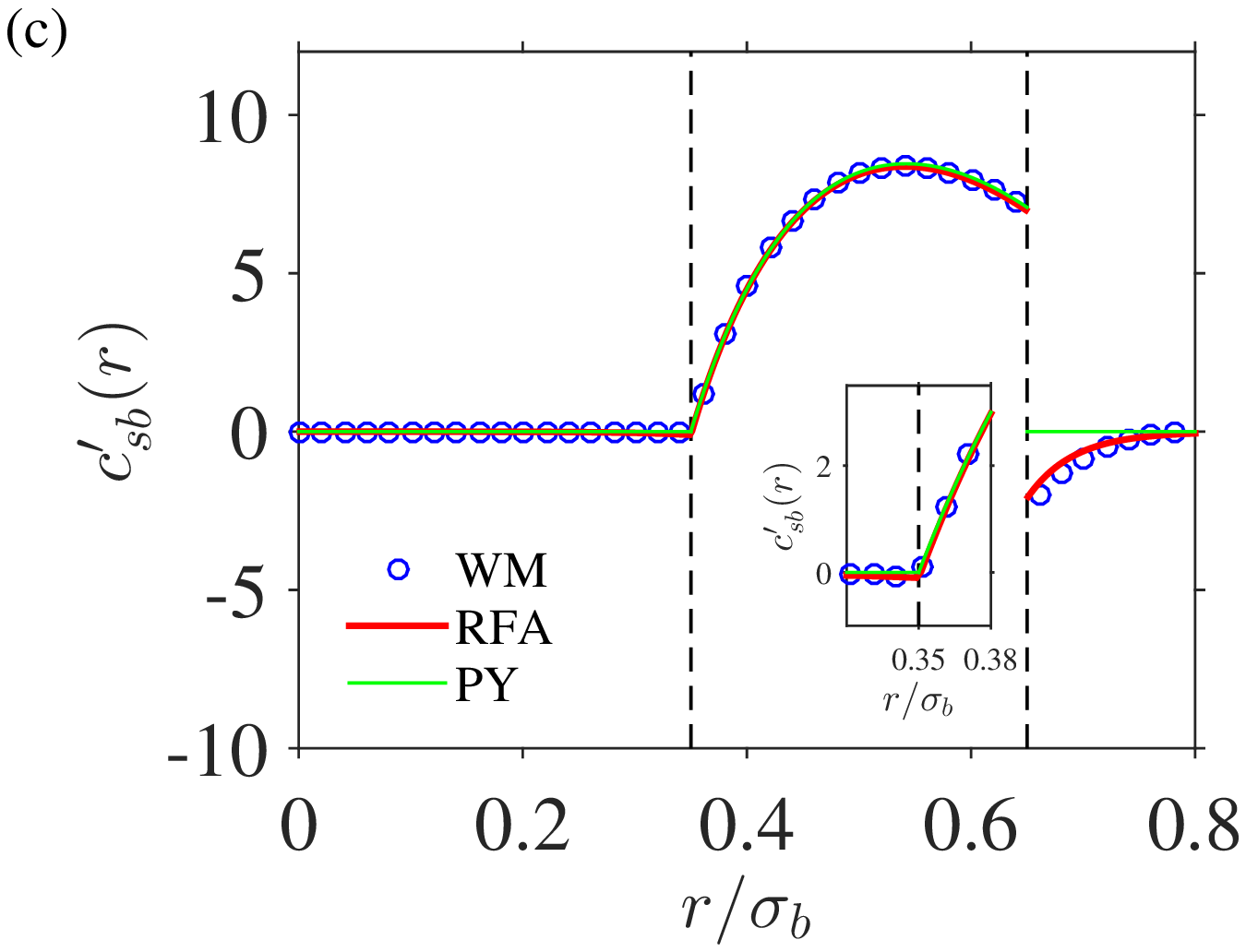}\\
\includegraphics[width=.65\columnwidth]{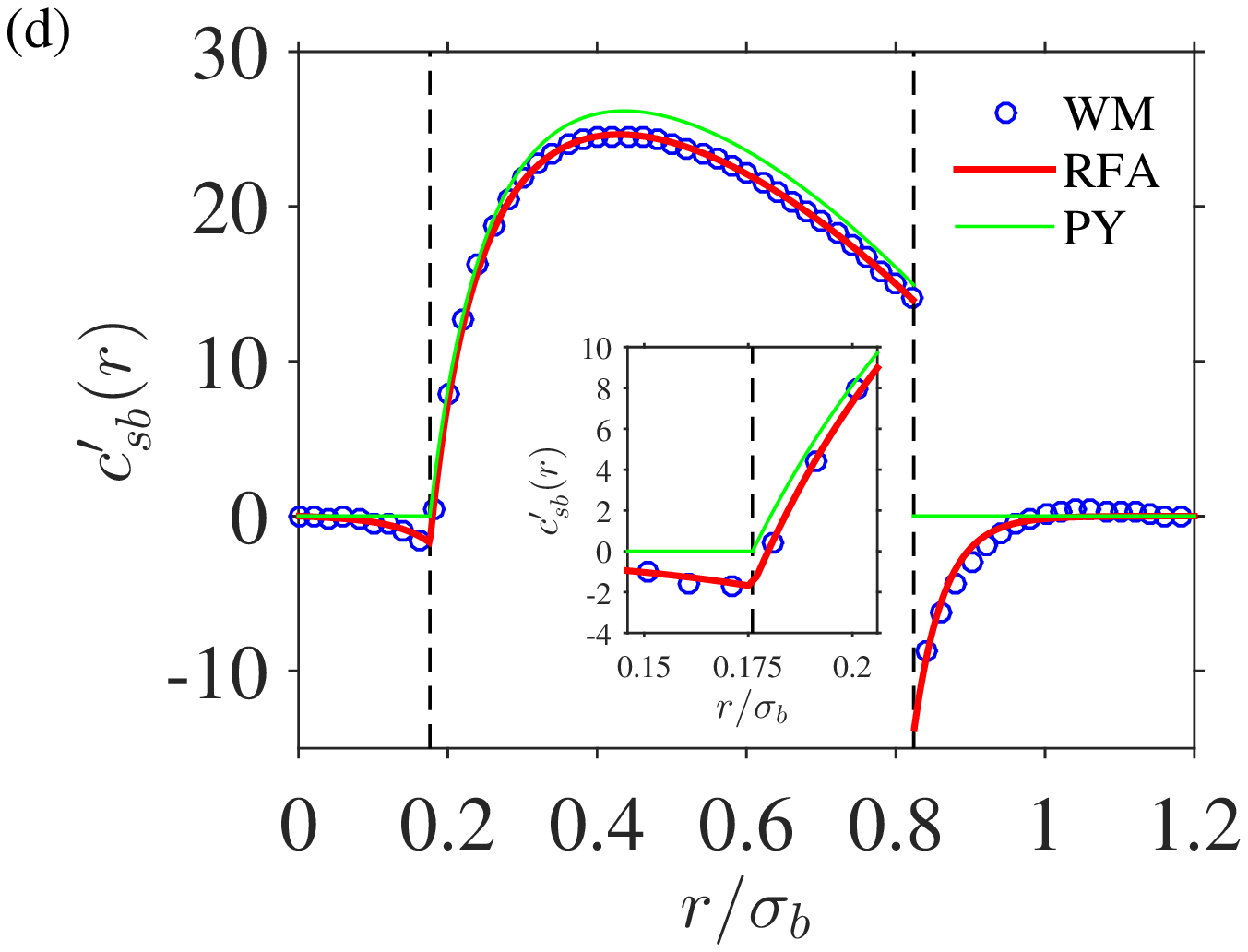}
\includegraphics[width=.63\columnwidth]{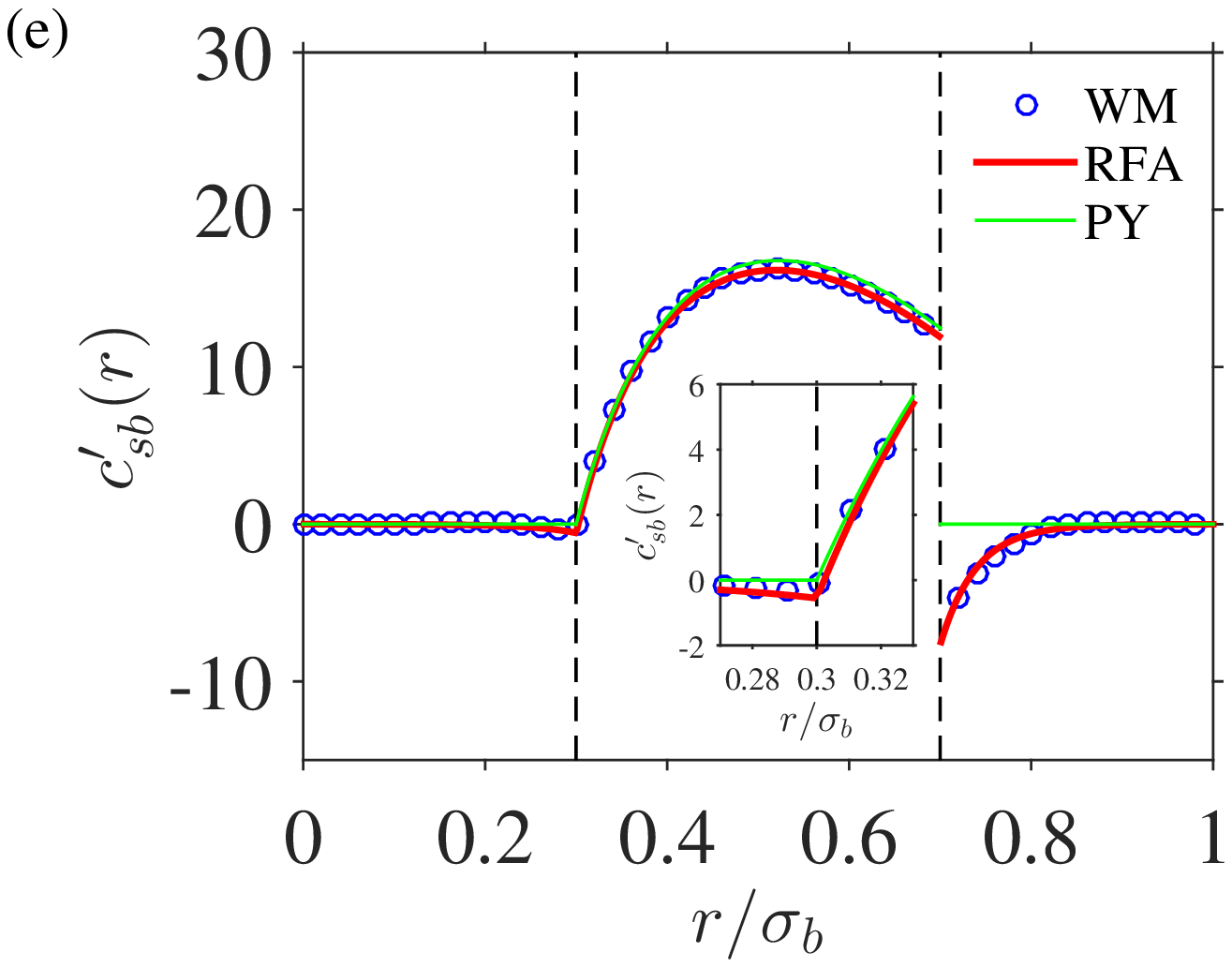}
\includegraphics[width=.65\columnwidth]{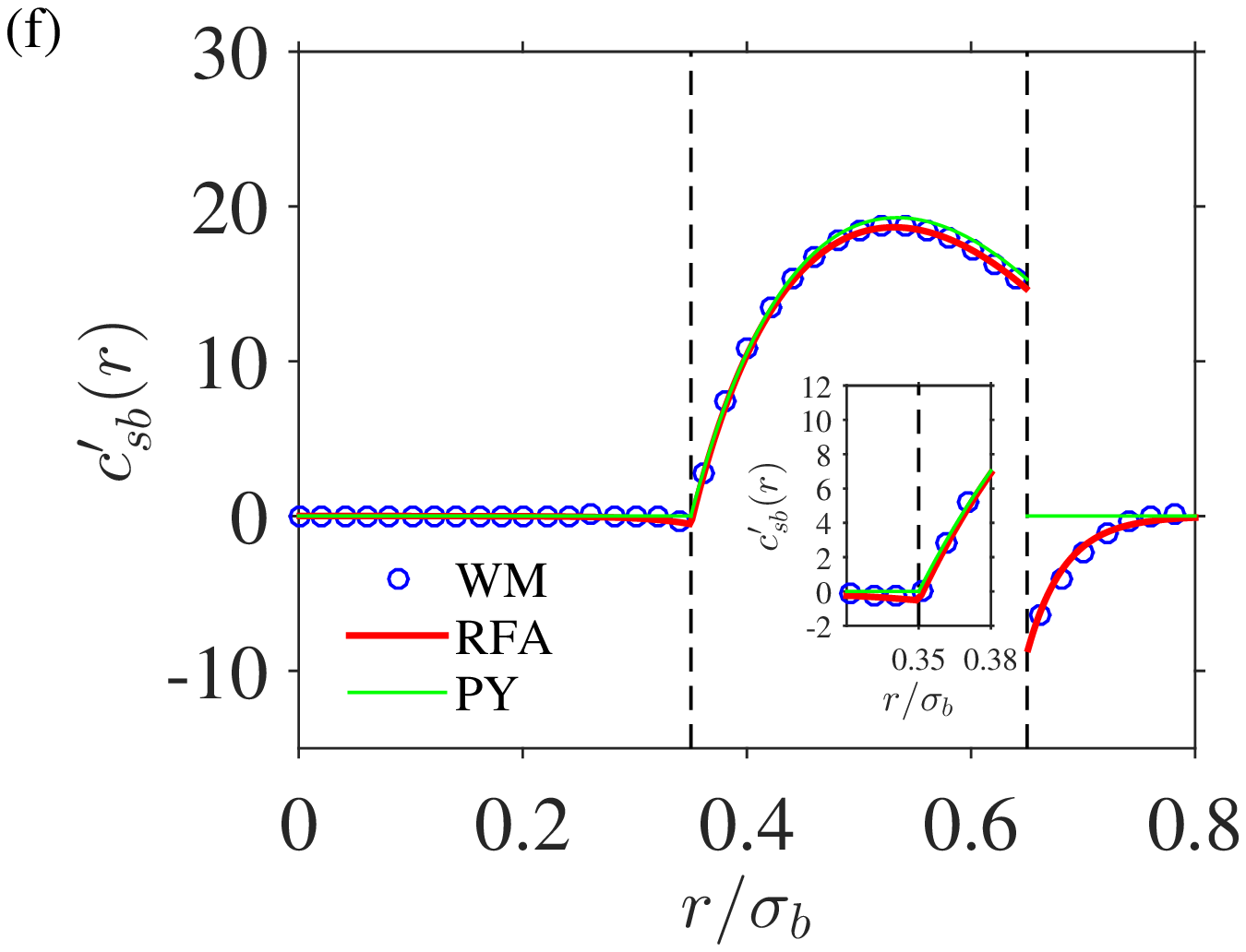}
\caption{Plot of  $c_{sb}'(r)$  for (a) $q=0.648$, $\eta_b=0.1$, $\eta_s=0.2$,  (b) $q=0.4$, $\eta_b=0.05$, $\eta_s=0.15$,  (c) $q=0.3$, $\eta_b=0.05$, $\eta_s=0.15$,  (d) $q=0.648$, $\eta_b=0.2$, $\eta_s=0.2$, (e) $q=0.4$, $\eta_b=0.2$, $\eta_s=0.15$, and (f) $q=0.3$, $\eta_b=0.2$, $\eta_s=0.15$. The blue circles are the WM results, the red thick lines correspond to the RFA values, and the green lines are PY. In each panel, the  vertical dashed lines signals the locations of $\lambda_{sb}$ and $\sigma_{sb}$. The insets show magnifications of $c_{sb}'(r)$ around $r=\lambda_{sb}$.}
\label{fig5}
\end{figure*}

\begin{figure*}
\includegraphics[width=.65\columnwidth]{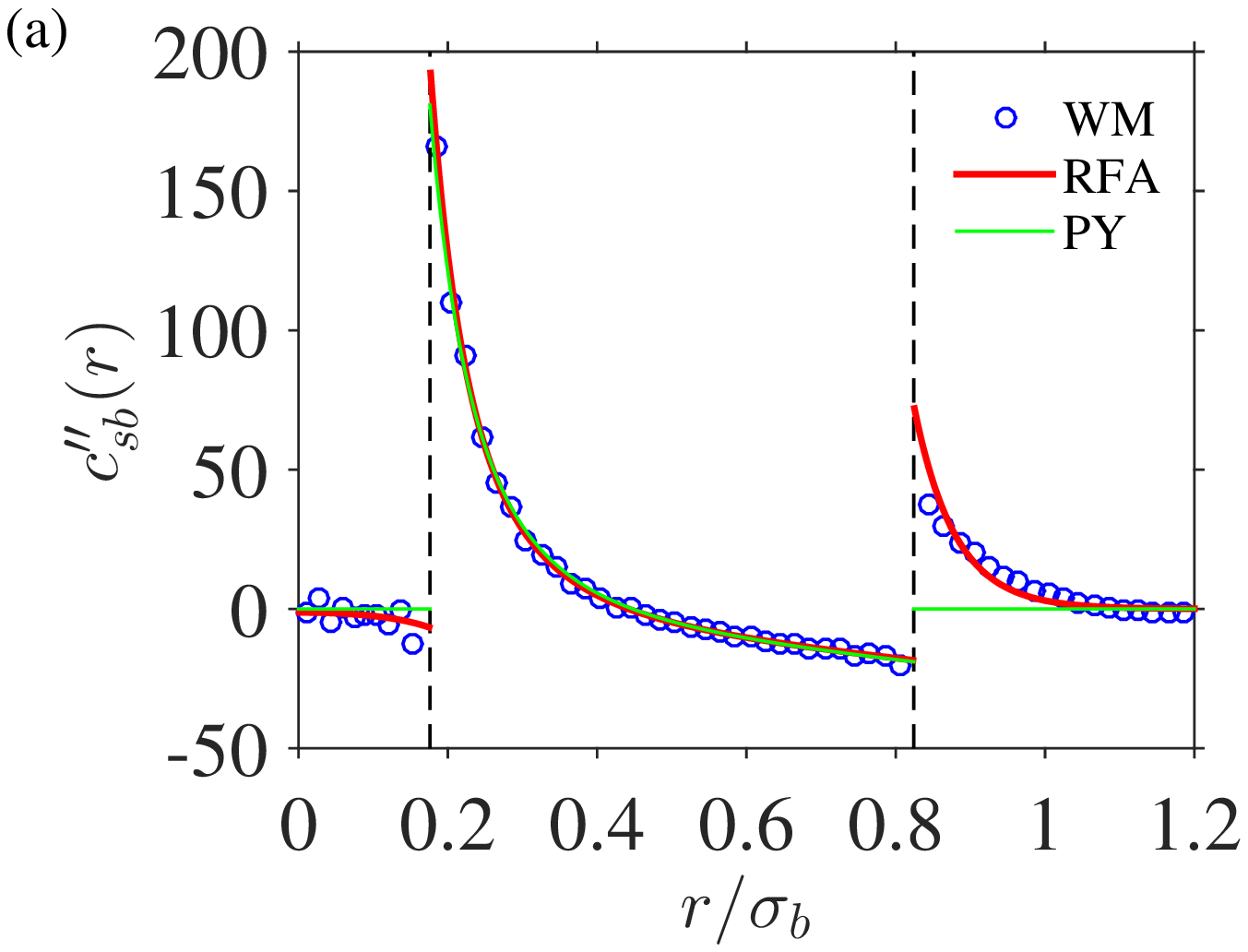}
\includegraphics[width=.63\columnwidth]{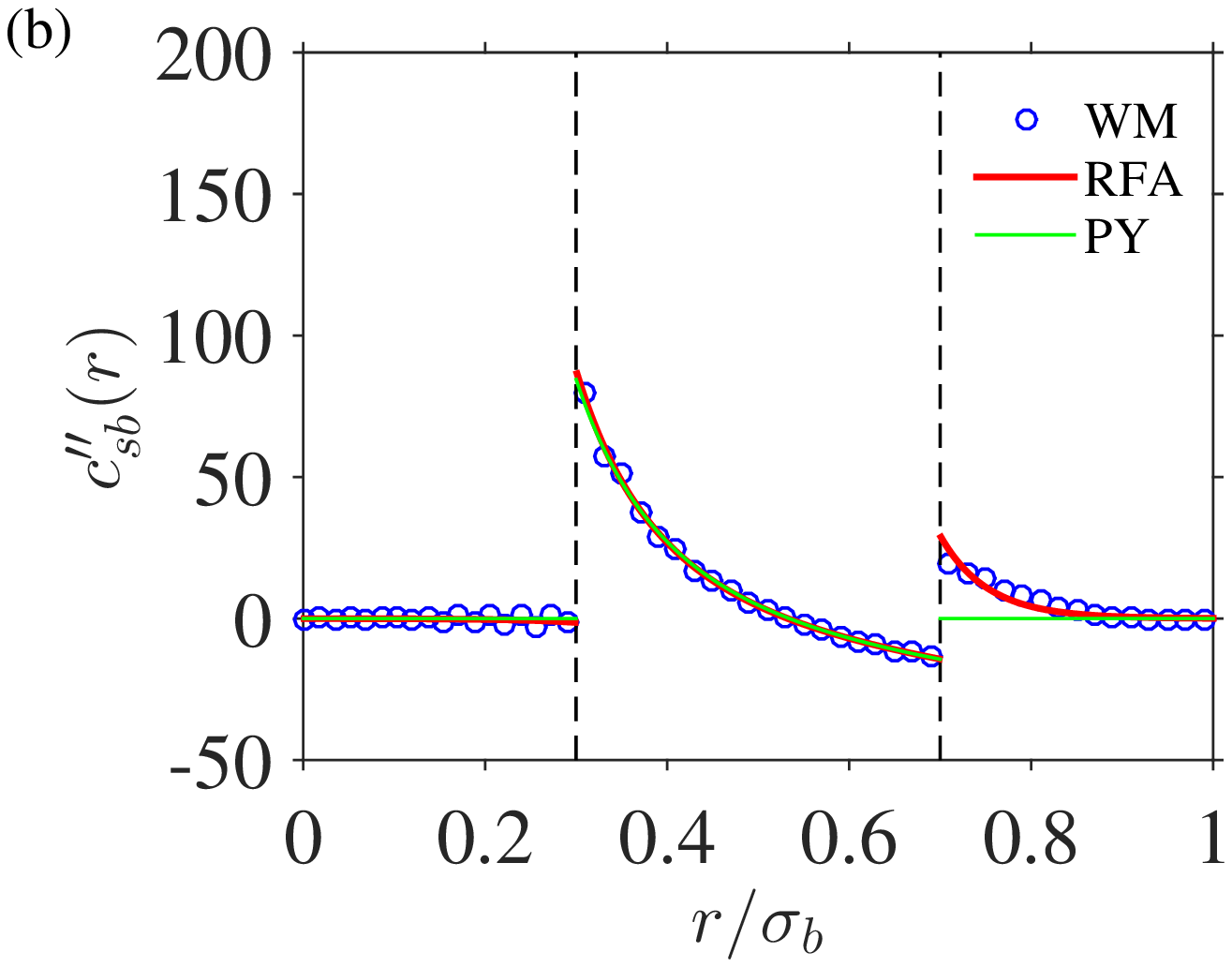}
\includegraphics[width=.65\columnwidth]{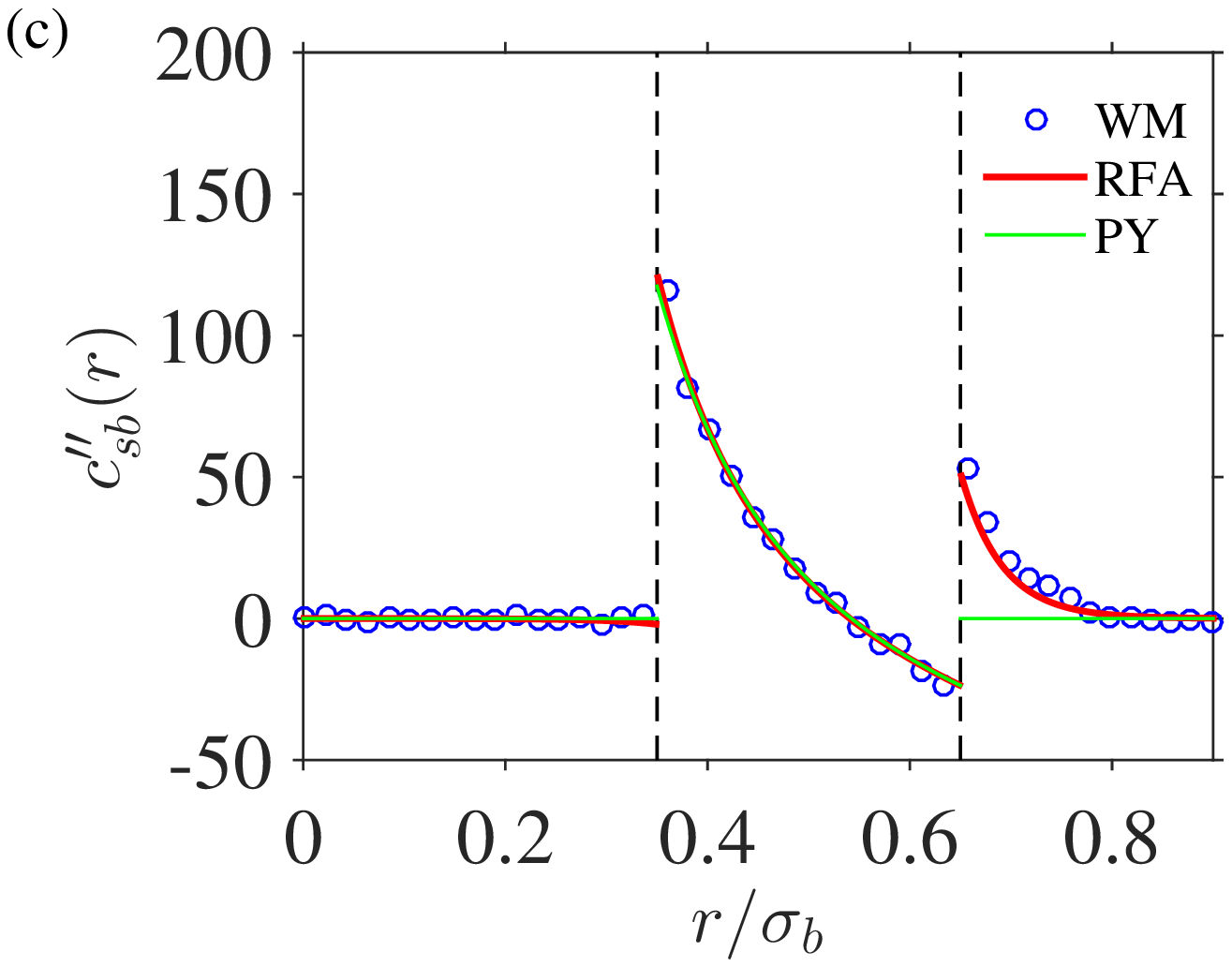}\\
\includegraphics[width=.65\columnwidth]{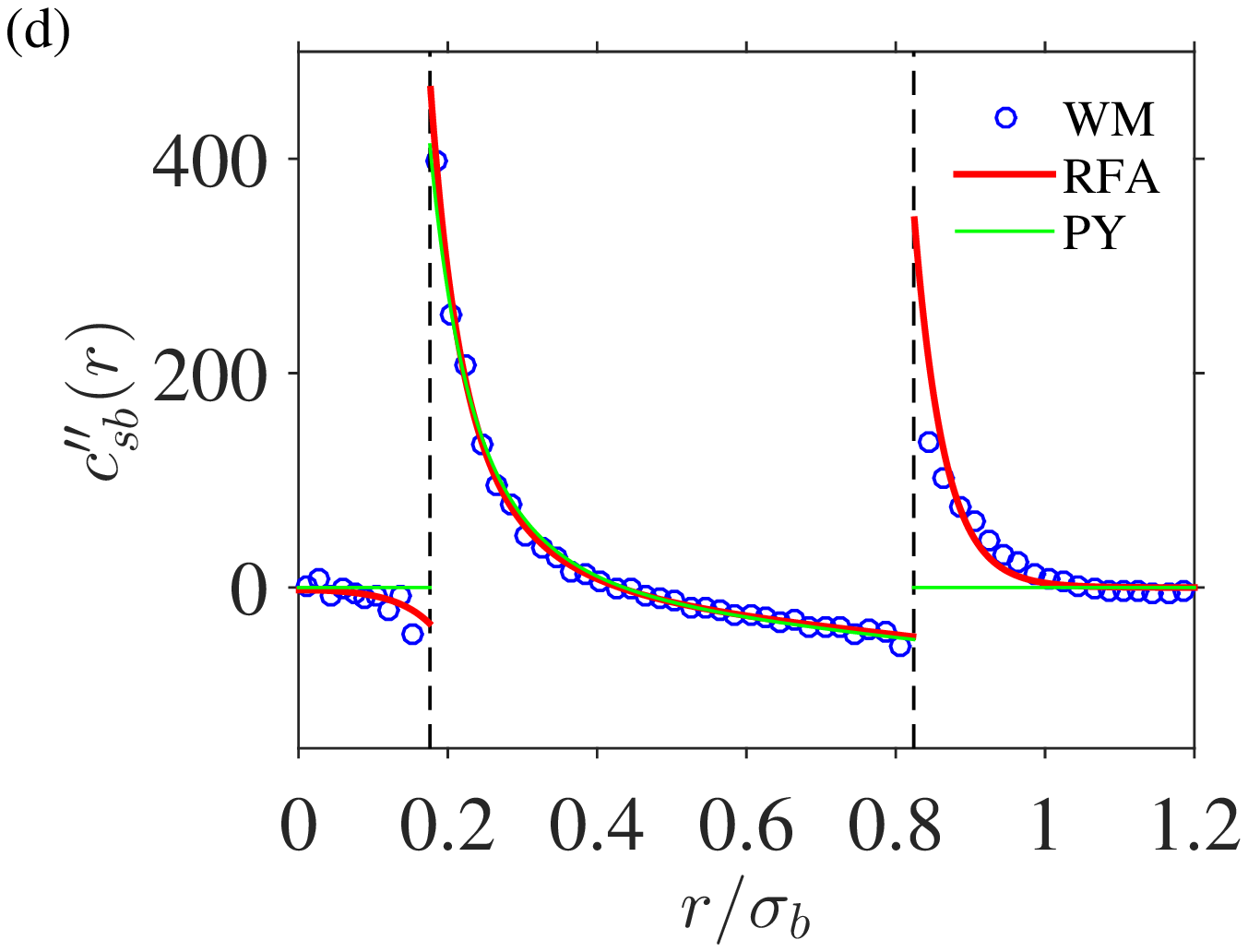}
\includegraphics[width=.63\columnwidth]{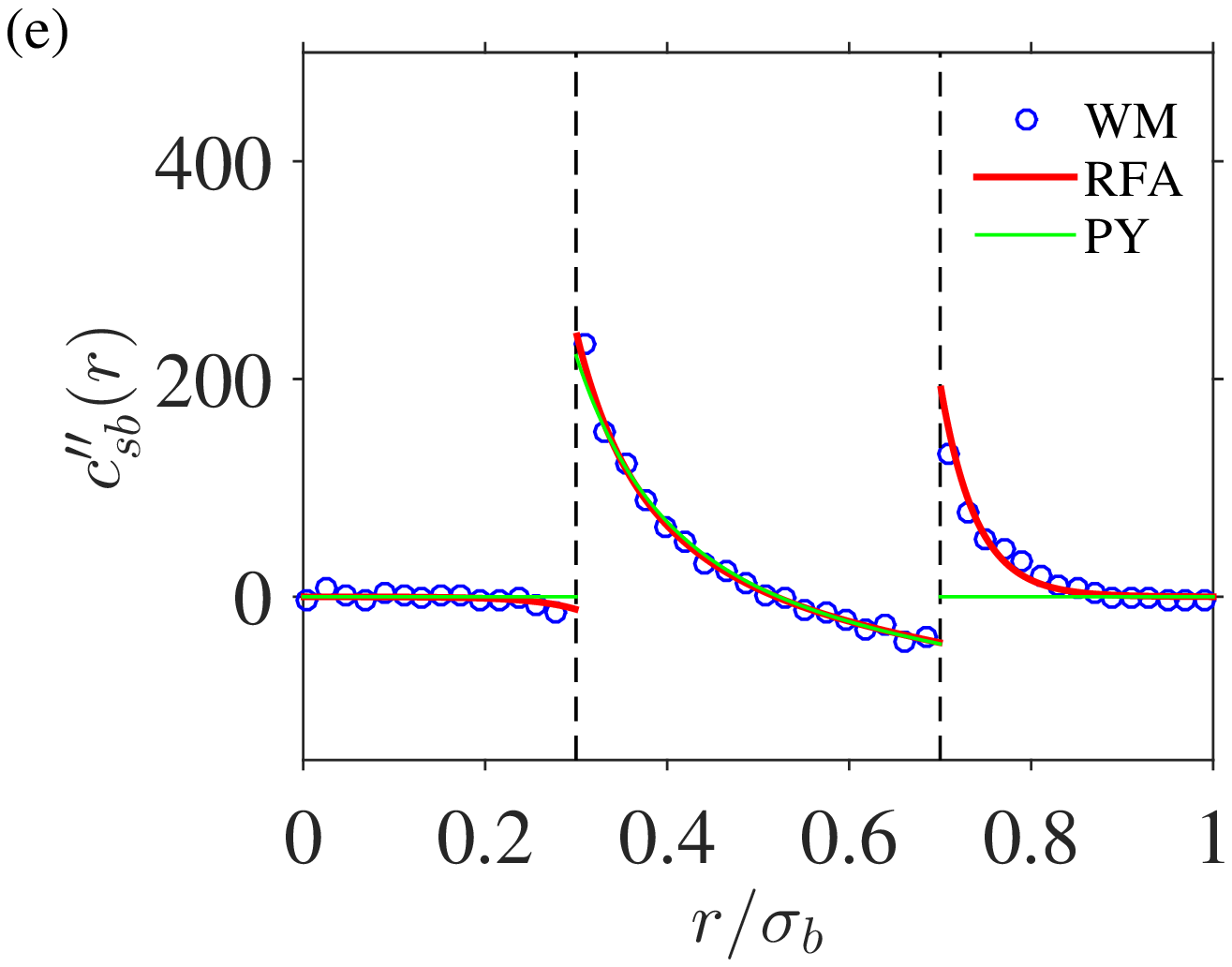}
\includegraphics[width=.65\columnwidth]{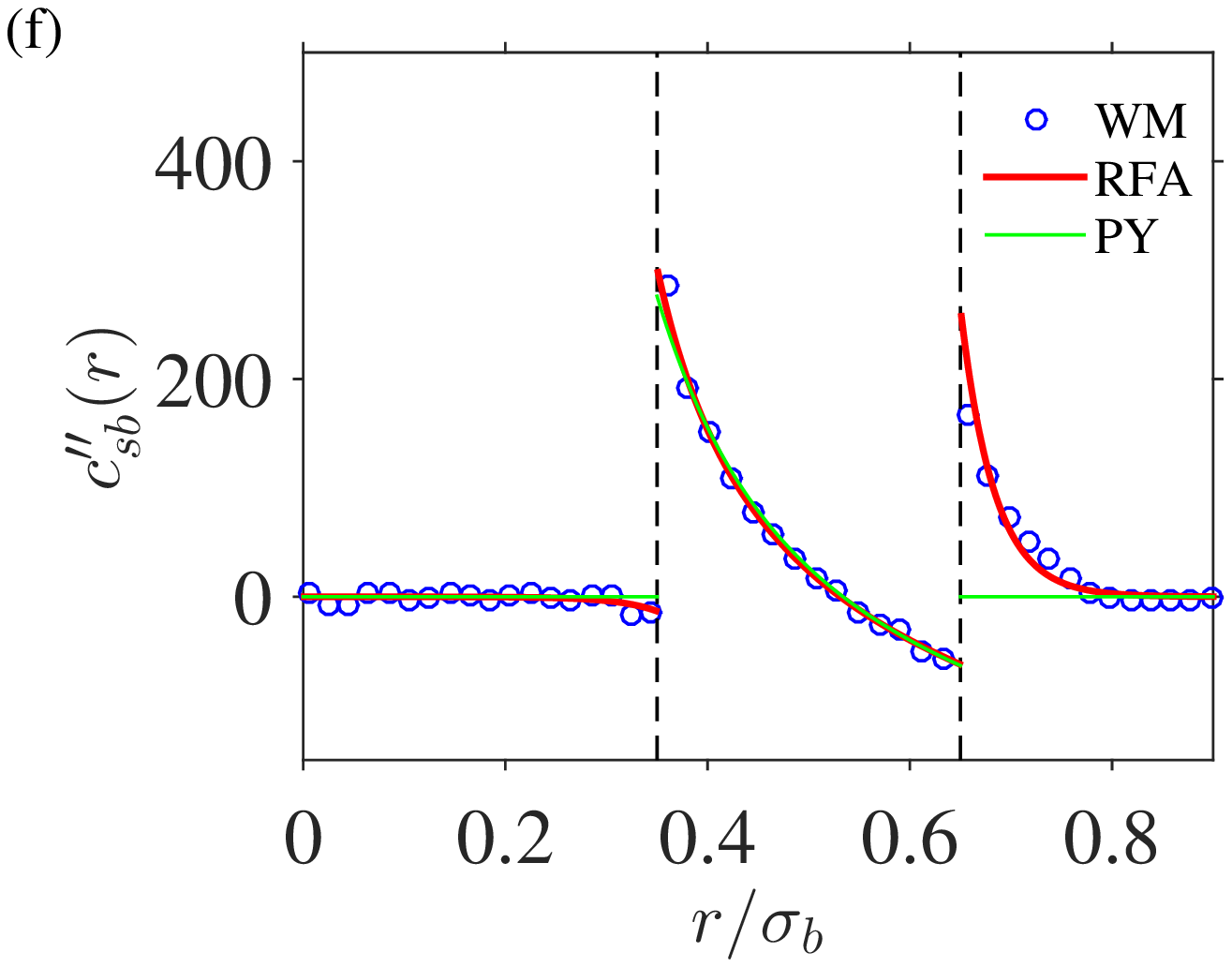}
\caption{Plot of $c_{sb}''(r)$  for (a) $q=0.648$, $\eta_b=0.1$, $\eta_s=0.2$,  (b) $q=0.4$, $\eta_b=0.05$, $\eta_s=0.15$,  (c) $q=0.3$, $\eta_b=0.05$, $\eta_s=0.15$,  (d) $q=0.648$, $\eta_b=0.2$, $\eta_s=0.2$, (e) $q=0.4$, $\eta_b=0.2$, $\eta_s=0.15$, and (f) $q=0.3$, $\eta_b=0.2$, $\eta_s=0.15$. The blue circles are the WM results, the red thick lines correspond to the RFA values, and the green lines are PY. In each panel, the vertical dashed lines signals the locations of $\lambda_{sb}$ and $\sigma_{sb}$.}
\label{fig6}
\end{figure*}

\begin{figure}
	\includegraphics[width=0.8\columnwidth]{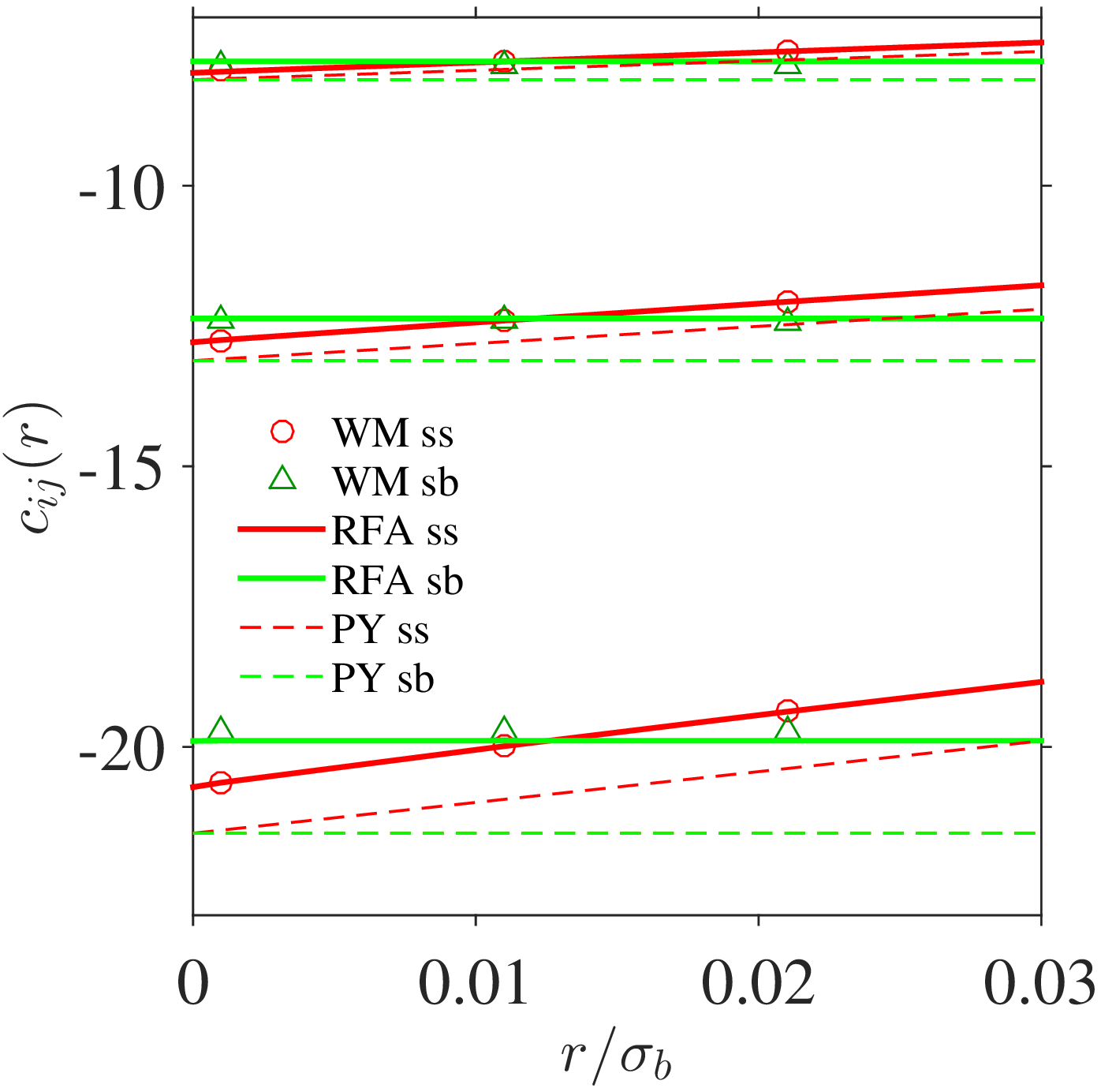}
	\caption{Plot of $c_{ss}(r)$ (red lines and symbols) and $c_{sb}(r)$ (green lines and symbols) near the origin for a size ratio $q=0.4$, a partial packing fraction $\eta_b=0.2$, and (from top to bottom) partial packing fractions $\eta_s=0.15$, $0.20$, and $0.25$. The symbols (red circles for $c_{ss}$ and green triangles for $c_{sb}$) are the WM results, the solid thick lines correspond to the RFA values, and the dashed lines represent the PY values.}
	\label{fig9}
\end{figure}

We next focus on the analysis of the cross function $c_{sb}(r)$. Careful inspection shows that, in all studied cases with both the WM method and the RFA, this function  has a minimum at some  $r=r_{sb}^{\min} < \sigma_{sb}$ (i.e., inside the core).
To the best of our knowledge this feature has not been so far discussed in the literature. Whether a physical meaning to such a feature may be ascribed or whether it may influence or be correlated with other physical properties is not clear to us at this stage, but we are persuaded that it should be further explored in the future.\\
%
%%%%%%%%%%%%%%%%%%%%%%%   Fig 2
\indent The characteristic form of the DCF $c_{sb} (r)$  inside the core ($r<\sigma_{sb}$) is shown in Fig.\ \ref{fig2}  for the size ratio $q=0.4$ and the partial packing fractions $\eta_b=0.2$ and $\eta_s=0.05$.
The inset in Fig.\ \ref{fig2} demonstrates that, in contrast to the PY theory, a  minimum value  at $r=r_{sb}^{\min}\approx \lambda_{sb}$ is present in the WM and RFA results. Nevertheless, the minimum $c_{sb}^{\min}\equiv c_{sb}(r_{sb}^{\min})$ is very shallow, and one may not notice it on a usual scale with typically obtainable accuracy. In fact, the general shape of the DCF $c_{sb}(r)$ inside the core is rather similar to the one of the  PY theory, and thus the minimum may be easily overlooked. It is worth noting that the fact that $c_{sb} (r)=\text{const}$ for $0<r<\lambda_{sb}$ in the PY theory may be linked to the tail property $c_{ij}(r)=0$ for $r>\sigma_{ij}$ in that approximation.

%%%%%%%%%%%%%%%%%  FIG.3
We have observed that the position of the minimum is always localized  close to  and above  $r=\lambda_{sb}=\frac{1-q}{2}\sigma_b$, its precise value slightly depending on the mixture composition and density. This is illustrated in Fig.\ \ref{fig2a}, where  $c_{sb}(r)$ in a spatial region around $r=\lambda_{sb}$ is plotted for three different  values of the size ratio $q$ and, in each case, two pairs (representing moderate and dense systems) of packing fractions $\eta_s$, $\eta_b$. Also, the corresponding PY results are plotted for a comparison.

%%%%%%%%%%%%%%%%   Fig.4
A more quantitative dependence of the minimum value ($c_{sb}^{\min}$) and its position ($r_{sb}^{\min}$)  on $\eta_s$ is presented in Fig.\ \ref{fig3} for $q=0.648$, $0.4$, and $0.3$ at some representative values of $\eta_b$. We observe that the value of the minimum becomes monotonically more negative as $\eta_s$ increases, with an excellent agreement between WM and RFA.
Also, we observe that, in the log-linear scale, this dependence is well represented by a linear function.
At a given $q$, the influence of density on the position $r_{sb}^{\min}$ is rather weak. Depending on the values of $q$ and $\eta_b$, the change of $r_{sb}^{\min}$ with increasing $\eta_s$ can be monotonic or nonmonotonic.\\
%
%%%%%%%%%%%%  Fig.5
\indent A more complete 3D view of the density dependence of $c_{sb}^{\min}$ and $r_{sb}^{\min}$ on density, as predicted by the RFA, is given by Fig.\ \ref{fig4} for the same values of $q$ as in Fig.\ \ref{fig3}.
One can observe that, in all the cases, $\Delta_{sb}\equiv r^{\min}_{sb}-\lambda_{sb}>0$ but $\Delta_{sb}/\sigma_b\sim 10^{-3}$. Furthermore, as density ($\eta_s$ and/or $\eta_b$) decreases, the difference  $\Delta_{sb}$ tends to $0$. The qualitative shape of the surface $r_{sb}^{\min}(\eta_s,\eta_b)$ is rather similar for different values of $q$: at  fixed $\eta_s$ (or $\eta_b$), $r_{sb}^{\min}$  first tends to increase and then to decrease with increasing $\eta_b$ (or $\eta_s$). So far as $c_{sb}^{\min}$, as already observed in Fig.\ \ref{fig3}, it decreases almost exponentially with increasing $\eta_s$ at fixed $\eta_b$.

Once the behavior of the DCFs has been discussed, let us consider  their first and second derivatives $c_{ij}'(r)$ and $c_{ij}''(r)$, respectively. Both in the WM and RFA schemes,
they can be obtained from the analytic knowledge of $\tilde{c}_{ij}(k)$ by application of expressions analogous to Eqs.\ \eqref{cijr} and \eqref{Eq:crN-crA}, except for the formal replacements
\begin{subequations}
\bal
\frac{\sin(kr)}{kr}\to &\frac{kr \cos(kr)-\sin(kr)}{kr^2},\\
\frac{\sin(kr)}{kr}\to &\frac{(2-k^2 r^2)\sin(kr)-2kr\cos(kr)}{kr^3},
\eal
\end{subequations}
for $c_{ij}'(r)$ and $c_{ij}''(r)$, respectively.

%%%%%%%%%%%%   Fig.6,7
The shapes of $c_{sb}'(r)$ and $c_{sb}''(r)$ are presented in Figs.\ \ref{fig5} and \ref{fig6}, respectively, for the same cases as in Fig.\ \ref{fig2a}. An excellent agreement between the WM and RFA values is again observed.
Moreover, the three approaches (WM, RFA, and PY) provide almost indistinguishable values of the second derivative $c_{sb}''(r)$ in the range $\lambda_{sb}<r<\sigma_{sb}$.
Note also that $c_{sb}'(r)$ and $c_{sb}''(r)$ are discontinuous at $r=\sigma_{sb}$ \cite{PBYSH20}, which is not surprising, given the fact that the DCFs $c_{ij}(r)$ themselves are discontinuous at $r=\sigma_{ij}$.
More interesting is the discontinuity of the second derivative $c_{sb}''(r)$ at $r=\lambda_{sb}$, its existence already captured by the PY theory \cite{L64}, according to which
\begin{align}
\label{Deltac''PY}
\Delta c_{sb}''(\lambda_{sb})\equiv &c_{sb}''(\lambda_{sb}^+)-c_{sb}''(\lambda_{sb}^-)\nn
  =&12\displaystyle{\frac{\sigma_{sb}}{\lambda_{sb}}}g_{sb}^c\left(\frac{\eta_s}{\sigma_s^2}g_{ss}^c+\frac{\eta_b}{\sigma_b^2}g_{bb}^c\right).
\end{align}
Taking into account that the PY values of $g_{ij}^c$ are exact to first order in density \cite{S16}, it follows that the discontinuity of $c_{sb}''(r)$ at $r=\lambda_{sb}$ is an exact property and not an artifact of the PY, RFA, or WM approaches. In fact, taking into account that \cite{L64}
\begin{subequations}
\begin{equation}
g_{ss}^c=1+\frac{5}{2}\eta-\frac{3}{2}\eta_b\left(1-\frac{\sigma_s}{\sigma_b}\right)+O(\rho^2),
\end{equation}
\begin{equation}
g_{bb}^c=1+\frac{5}{2}\eta+\frac{3}{2}\eta_s\left(\frac{\sigma_b}{\sigma_s}-1\right)+O(\rho^2),
\end{equation}
\begin{equation}
g_{sb}^c=1+\frac{5}{2}\eta+\frac{3}{2}\frac{\lambda_{sb}}{\sigma_{sb}}\left(\eta_s-\eta_b\right)+O(\rho^2),
\end{equation}
\end{subequations}
one gets the exact result
\begin{align}
\Delta c_{sb}''(\lambda_{sb})=&12\left[\frac{\sigma_{sb}}{\lambda_{sb}}\left(\frac{\eta_s}{\sigma_s^2}+\frac{\eta_b}{\sigma_b^2}\right)(1+5\eta)\right.\nn
&\left.
+\frac{3}{2}\left(\frac{\eta_s^2}{\sigma_s^2}-\frac{\eta_b^2}{\sigma_b^2}\right)-12\eta_s\eta_b\frac{\lambda_{sb}\sigma_{sb}}{\sigma_s^2\sigma_b^2}\right]+O(\rho^3).
\end{align}

Since the singularities of $c_{ij}(r)$  are independent of $Q$ in Eqs.\ \eqref{Eq:crN-crA} and $c_{sb}^{\text{num}}(r)$ is regular, it turns out that the singularities of $c^{\text{tail}}_{ij}(r)$ determine those of the full functions $c_{ij}(r)$. In particular, the discontinuity of the second derivative $c_{sb}''(r)$ at $r=\lambda_{sb}$ is
\beq
\label{Deltac''}
\Delta c_{sb}''(\lambda_{sb})=-\frac{ K_{sb}}{4\pi \lambda_{sb}},
\eeq
where $K_{sb}$ is the coefficient of a term of the form $\cos (k\lambda_{sb})$ in the function $\tilde{c}_{sb}^{(4)}(k)$ defined in Eq.\ \eqref{cijkn}.
Taking into account that $\cos(k\sigma_{sb})\cos(k\sigma_{s,b})=\frac{1}{2}\left[\cos(k\lambda_{sb})+\cos(k(\sigma_{sb}+\sigma_{s,b}))\right]$ one can find from Eq.\ (16b) of Ref.\ \cite{PBYSH20} that  $K_{sb}=-\frac{1}{2}C_{sb}^{(1)}\left[ \rho_s C_{ss}^{(1)} + \rho_b C_{bb}^{(1)} \right]$, where $C_{ij}^ {(1)}=4\pi\sigma_{ij}g_{ij}^c$ \cite{PBYSH20}. Inserting all of this into Eq.\ \eqref{Deltac''}, one finally arrives at Eq.\ \eqref{Deltac''PY}. This proves that the relationship between $\Delta c_{sb}''(\lambda_{sb})$ and the contact values $g_{ij}^c$ given by Eq.\ \eqref{Deltac''PY} is an exact property, even though the PY contact values are only approximate.

A peculiar prediction of the PY theory is that the zero-separation values of $c_{ss}(r)$ and $c_{sb}(r)$ are equal, i.e., $c_{ss}^\py(0)=c_{sb}^\py(0)$. However, as Fig.\ \ref{fig9} shows, this simple property  is not  fulfilled by either WM or RFA, and one actually has $c_{sb}(0)>c_{ss}(0)$, the difference  $c_{sb}(0)-c_{ss}(0)$  tending to increase as the packing fraction of the small spheres increases.

\section{Concluding remarks}
\label{sec4}
In this work, we have confirmed the excellent performance of the RFA for additive BHS mixtures when compared with the simulation-fed WM scheme \cite{PBYSH20,PYSHB21}, this time in connection with the DCFs.\\
\indent We have mainly focused on the properties of the cross DCF $c_{sb}(r)$ and highlighted a new feature of this structural function. Such function (for all BHS fluids) has a minimum inside the core, its location and magnitude  depending on density and mixture composition. However, the minimum is rather shallow and hence the nonmonotonic character of the DCF $c_{sb}(r)$ may be hardly visible. The minimum is always localized near $r=\lambda_{sb}$, and we have been able to analyze some characteristic dependence of both its value and position on density and composition.
The observed disappearance of the minimum at the low density limit is in agreement with the known zero-density limit of $c_{ij}(r)$ (negative of the Mayer $f$ functions).\\
\indent The physical origin of such a minimum and its relation to other properties are not clear at this stage. The comparison with the PY result may suggest that oversimplification of the $c_{sb}(r)$ outside the core [i.e., the condition $c_{sb}^\py(r>\sigma_{sb})=0$]  may lead to the monotonic behavior of $c_{sb}^\py(r)$ inside the core in this approximation.\\
\indent Moreover, we have studied the behavior of the first and second spatial derivatives $c_{sb}'(r)$ and $c_{sb}''(r)$, respectively. From the analysis of  $c_{sb}''(r)$ one concludes is that there is a discontinuity of this derivative at $r=\lambda_{sb}$ whose size has exactly the same expression in terms of the contact values $g_{ij}^c$ as that of the corresponding PY result.\\
\indent We hope that our study can stimulate further investigations on the properties of the DCFs in fluid mixtures different from the additive BHS model.

\begin{acknowledgments}
S.B.Y. and A.S. acknowledge financial support from Grant PID2020-112936GB-I00 funded by MCIN/AEI/10.13039/501100011033, and from Grants IB20079 and GR18079 funded by Junta de Extremadura (Spain) and by ERDF A way of making Europe. It must be acknowledged that the Grant GR18079 also financed the summer visit of  M.L.H. to Universidad de Extremadura, where a first draft of the paper was prepared.  S.P. is grateful to the  Universidad de Extremadura, where most of this work was carried out during his scientific internship, which was supported by Grant No. DEC-2018/02/X/ST3/03122 financed by the National Science Center, Poland. Some of the calculations were performed at the Pozna\'{n} Supercomputing and Networking
Center (PCSS).
\end{acknowledgments}

%\bibliographystyle{apsrev}

%\bibliography{C:/AA_D/Dropbox/Mis_Dropcumentos/bib_files/liquid}

%apsrev4-2.bst 2019-01-14 (MD) hand-edited version of apsrev4-1.bst
%Control: key (0)
%Control: author (8) initials jnrlst
%Control: editor formatted (1) identically to author
%Control: production of article title (0) allowed
%Control: page (0) single
%Control: year (1) truncated
%Control: production of eprint (0) enabled
%

\end{document}